\documentclass[amsfonts,amssymb,twocolumn,showpacs]{revtex4}
\usepackage{mathrsfs}
\usepackage{graphicx}

\begin{document}

\title{Relativistic quantum effects of Dirac particles simulated by ultracold atoms}

\author{Dan-Wei Zhang$^{1,2}$, Z. D. Wang$^2$, Shi-Liang Zhu$^{1,}$}
\altaffiliation{Electronic address: slzhu@scnu.edu.cn}

\affiliation{$^1$Laboratory of Quantum Information Technology,
School of Physics and Telecommunication and Engineering, \\South
China Normal University, Guangzhou, China\\
$^2$Department of Physics and Center of Theoretical and
Computational Physics, \\The University of Hong Kong, Pokfulam Road,
Hong Kong, China}

\date{\today}

\begin{abstract}
Quantum simulation is a powerful tool to study a variety of problems
in physics, ranging from high-energy physics to condensed-matter
physics. In this article, we review the recent theoretical and
experimental progress in quantum simulation of Dirac equation with
tunable parameters by using ultracold neutral atoms trapped in
optical lattices or subject to light-induced synthetic gauge fields.
The effective theories for the quasiparticles  become relativistic
under certain conditions in these systems, making them ideal
platforms for studying the exotic relativistic effects. We focus on
the realization of one, two, and three dimensional Dirac equations
as well as the detection of some relativistic effects, including
particularly the well-known {\sl Zitterbewegung} effect and Klein
tunneling. The realization of quantum anomalous Hall effects  is
also briefly discussed.

\vspace{5mm} \noindent {\bf Keywords:} ultracold atoms, Dirac
equation, quantum simulation

\end{abstract}

\maketitle

\tableofcontents

\section{Introduction}

\noindent As first pinpointed by Richard Feynman in 1982, a quantum
computer could be used to simulate a quantum system efficiently
without experiencing an exponential explosion and slowdown, which a
classic computer would presumably encounter when simulating quantum
phenomena~[1]. Although quantum computers are not yet available, one
can still hope to create analog quantum simulators which are
designed to have the same Hamiltonian as another system of interest.
The recent rapid progress in quantum coherent control~[2] of neutral
atoms, photons, or ions in some systems makes them become ideal
quantum simulators~[3], which include neutral atoms in optical
lattices (OLs), arrays of cavities, trapped ions, quantum dots,
superconducting circuits, nuclear magnetic resonance (NMR) and so
on.

Analog quantum simulators would be able to provide highly
controllable platforms to study difficult problems in various
fields, ranging from high-energy physics, condensed-matter physics
to quantum chemistry. Some recent notable achievements in quantum
simulation are listed as follows: addressing neutral atoms in OLs to
simulate the Bose-Hubbard model with the quantum phase transition
from a superfluid to a Mott insulator~[4-6], as well as to simulate
the antiferromagnetic spin chains~[7]; simulating a quantum
magnet~[8], frustrated Ising model~[9], relativistic
dynamics~[10,11], and open-system dynamics~[12] with trapped ions;
implementing a hydrogen molecule quantum simulation with NMR~[13].
Besides, many other experimental proposals of quantum simulation
with different simulators are suggested (see the review~[3] and the
references therein).

One of the most promising quantum simulators is the cold atom system
since the realization of condensation of bosonic atoms~[14,15] and
fermionic atom pairs~[16,17] has led to numerous advances with
applications in coherent manipulation of neutral atoms, such as
creating cold molecules in quantum chemistry (see the
reviews~[18,19] and the references therein). Ultracold neutral atoms
in OLs are quite suited for mimicking condensed-matter
physics~[20,21], and may shew new light on strongly correlated state
problems like high-temperature superconductivity~[22], fractional
quantum Hall effect~[23,24], and the quantum magnetism~[25].
Ultracold atomic gases in disordered optical potentials pave the way
towards the realization of versatile quantum simulators for the
investigations of Anderson localization~[26,27], which may help
solve some open questions relying on the interplay of disorder and
interactions~[28,29]. In addition, some phenomena and effects in
high-energy physics and cosmology can be mimicked, such as quantum
simulation of black holes~[30] and cosmic inflation~[31] by using
Bose-Einstein condensates (BECs), or superstrings~[32] and
supersymmetry~[33,34] by using Bose-Fermi mixtures.

Since the relativistic Dirac fermions were found in graphene
~[35,36], a substantial amount of efforts has been devoted to the
understanding of exotic relativistic effects in solid state systems
and the searching for other relativistic systems such as the recent
discovery of topological insulators~[37]. Inspired by those exciting
results and the state-of-the-art technologies in quantum control of
atoms, one specific topic among ultracold neutral atom simulators
has arisen, that is, quantum simulation of the Dirac equation and
the related relativistic effects.

The Dirac equation $i\hbar\partial_t \Psi=(c\vec{\alpha}\cdot{\bf
\vec{p}}+mc^2\beta)\Psi$, where $\vec{\alpha}$ and $\vec{\beta}$ are
the Dirac matrices, and $c$ is the speed of light, was first
proposed by Dirac who was seeking a relativistic and quantum
mechanical equation to describe spin $1/2$ particles with mass $m$.
This equation naturally provides a description of the electron spin,
which is an assumption in the Schr\"{o}dinger equation. More
surprisingly, it predicts the existence of antiparticles~[38], which
was soon confirmed by the experimental discovery of positrons. The
Dirac equation also predicts some exotic effects, and the most
famous ones are {\sl Zitterbewegung} (ZB)~[39], an unexpected
trembling motion of free relativistic particles, and Klein tunneling
(KT)~[40], which describes relativistic particles penetrating
through high and wide potential barriers without exponential damping
expected in non-relativistic tunneling processes. Despite these
relativistic effects have attracted lots of interest over years,
they are failed to be directly tested by elementary particles due to
currently unaccessible experimental techniques. Taking a free
electron as an example, the frequency and the amplitude of ZB is on
the order of $10^{21}$ Hz and $10^{-3}$ {\AA}, respectively, and the
electric field gradients requirement for observing KT is on the
order of $10^{18}$ V/m. All those conditions are out of reach in
current technologies.

However, we will see that ultracold atoms loaded in some optical
lattices or subject to certain synthetic gauge fields behave as the
relativistic particles under suitable conditions. Furthermore, cold
atom simulators offer us rather more degrees of freedom to control
the relativistic quasiparticles, such as the dimensions, the
effective mass and the effective speed of light. Associated with the
controllable atomic interactions and disorder, these systems provide
us an ideal platform to investigate the interesting relativistic
effects such as the never-before-seen ZB and KT for free particles.
Although the original Dirac equation is specific to fermions with
spin $1/2$, not bosons with an integer spin, we will see that one
can still simulate the Dirac equation with bosons: a pseudo-spin
$1/2$ can be introduced to the bonons and the dynamics of such
pseudo-spin should be described by the Dirac equation.

In this article, we review the recent theoretical investigation and
experimental proposals in quantum simulation of the Dirac equation
and some related effects by using ultracold neutral atoms. This
review is organized as follows. In Section II, we introduce the
realization and detection of the two and three-dimensional
relativistic quasiparticles by using cold atoms in OLs with a wide
range of structures. The realization of quantum anomalous Hall
effect (QAHE) is also briefly introduced. In Section III, we
introduce another kind of proposals for simulating the tunable Dirac
equation that are based on generating effective gauge fields on bulk
atomic gases. The simulated Dirac equations do not need OLs and thus
operates in a continuous regime. In Section IV, we show that one can
observe the well-known ZB and KT by using cold atoms with the
schemes reviewed in Section III. Furthermore, we show that an exotic
macroscopic KT described by the nonlinear Dirac equation (NLDE)
exhibits. Finally, in Section V, we give our summary and conclusion
of this review, together with some prospects on the quantum
simulation of relativistic effects and related fields in physics.

\section{Simulation of Dirac equation with ultracold atoms in OL systems}

\noindent In this section, we first review that ultracold neutral
fermionic atoms in a honeycomb (hexagonal) OL, similar with
electrons in graphene, can mimic 2D massless and massive Dirac
fermions~[41-47]. Then we demonstrate that, it is also possible to
simulate 2D Dirac fermions in a so-called $\mathcal{T}_3$
OL~[48,49], a line-centered-square (LCS) OL~[50], and even in a
simple square OL combing with a synthetic gauge potential~[51-58].
Furthermore, we show that 3D Dirac fermions may be realized with
cold atoms properly loaded in some cubic OLs~[59-61]. The
realization of quantum anomalous Hall effect (QAHE) associated to
the parity anomaly of 2D Dirac fermions is also briefly introduced.
In words, the emergent relativistic Dirac fermions in these OL
systems originate from the single-particle band dispersion, which
provides the Dirac cones in the Brillouin zone. Some lattices with
symmetric structures like hexagonal, ${\mathcal T}_3$ and LCS in the
real space, may directly support the needed structures of the
momentum space; the simple square lattice may fail, and thus needs
the aid of external fields. The near-half-filling condition allows
the Fermi level being close to the Dirac points, and in the long
wavelength limitation, the dynamics of the cold atoms are described
well by the relativistic theories. The OLs allow one to tune the
tunneling probabilities from site to site as well as the atomic
interactions, and may make the relativistic quasiparticles
realizable and controllable in these systems.

\vspace{3mm}
\includegraphics[width=7cm,height=6.5cm]{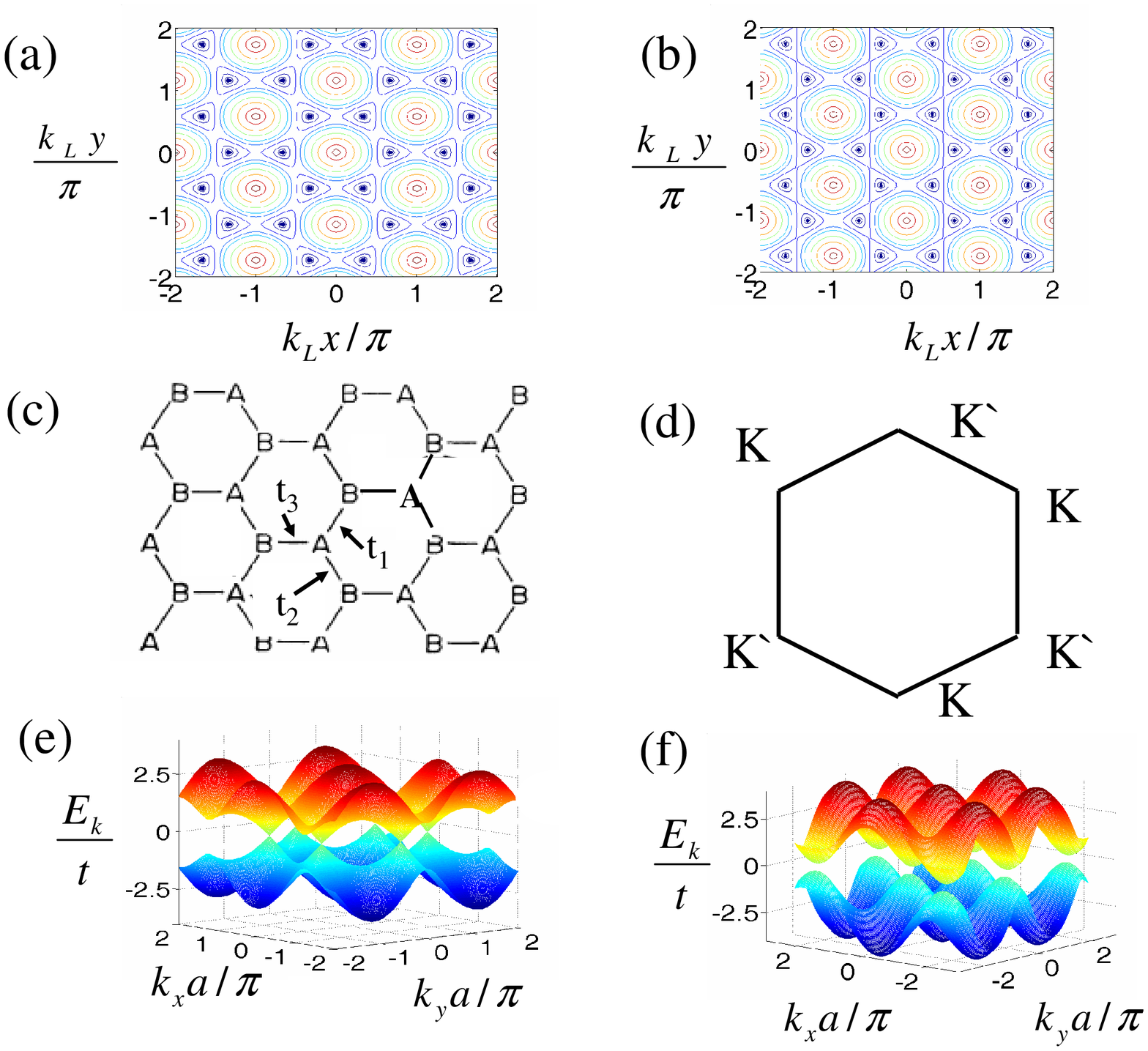} \vspace{1mm}
{\baselineskip
10.5pt\renewcommand{\baselinestretch}{1.05}\footnotesize

\noindent{\bf Fig.~1}\quad The honeycomb OLs. (a),(b) The contours
with three potentials described in Eq.~(1). The minima of the
potentials are denoted by the solid dots. All $V^0_j$ are the same
in (a), and $V^0_1=V^0_2=0.91V^0_3$ in (b). (c) Decomposition of the
hexagonal lattice as two triangular sublattices $A$ and $B$ with
anisotropic tunnelling. (d) The Brillouin zone of the hexagonal
lattice. The dispersion relations are shown in (e) for $\beta=1$
(gapless state) and (f) for $\beta=2.5$ (gapped state). Reproduced
from Ref.~[41], Copyright \copyright~2007 the American Physical
Society.

}\vspace{3mm}

\subsection{Simulation of 2D relativistic Dirac fermions}

\noindent The graphene material, formed with a single layer of
carbon atoms~[35,36] with its emergent massless Dirac fermions, has
recently attracted strong interest in condensed-matter physics. One
topic naturally arises is to mimic the graphene and the relativistic
quasiparticles with cold atom in a similar 2D hexagonal
lattice~[41-47]. In addition to the graphene-type lattice, it is
also interesting to search other proposals for quantum simulation of
2D Dirac equation with cold atoms in OLs of other structures
~[48-58].

\subsubsection{Mimic graphene: Cold atoms in a honeycomb OL}

\noindent Simulating Dirac equations with cold atoms loaded in a
honeycomb OL was proposed in the first time in Ref.~[41]. We review
this proposal in this section. Considering single-component
fermionic atoms (e.g., spin-polarized atoms $^{40}$K, $^{6}$Li,
etc.) in a two-dimensional ($x$-$y$ plane) hexagonal OL formed with
three detuned standing-wave lasers~[62]

\vspace{0.25cm}\noindent $\displaystyle V(x,y)=\sum_{j=1,2,3} V^0_j
\sin^2\Big[k_L(x \cos\theta_j+y
\sin\theta_j)+\frac{\pi}{2}\Big],$\hfill (1)

\vspace{0.25cm} \noindent where $\theta_1=\pi/3$, $\theta_2=2\pi/3$,
$\theta_3=0$, and $k_L$ is the optical wave vector. It is easy to
tune the potential barriers $V^0_j$ by varying the laser intensities
along different directions to form a standard hexagonal lattice for
$V^0_1=V^0_2=V^0_3$, and a hexagonal lattice but with a finite
anisotropy for different $V^0_j$ as shown in Fig.~1(a) and 1(b),
respectively. A hexagonal lattice consists of two sublattices
denoted by $A$ and $B$, as shown in Fig.~1(c). For single-component
fermionic atoms, the atomic collisions are negligible at low
temperatures. The tight-binding Hamiltonian of the system is then
given by

\vspace{0.25cm}$\displaystyle \mathcal{H}=-\sum_{\langle i,j
\rangle}t_{ij}(a^\dag_i b_j+{\rm H.c.}),$ \hfill (2)

\vspace{0.25cm} \noindent where $\langle i,j \rangle$ represents the
neighboring sites, $a_i$ and $b_j$ denote the fermionic mode
operators for the sublattices A and B, respectively. The tunneling
rates $t_{ij}$ depend on the tunneling directions in an anisotropic
hexagonal lattice, and we denote them as $t_1$, $t_2$, $t_3$
corresponding to the three different directions as shown in Fig.
1(c). For simplicity, we assume $t_1=t_2=t$ and $t_3=\beta t$ with
$\beta$ being the anisotropy parameter. As the atomic tunneling rate
in an OL is exponentially sensitive to the potential barrier, this
control provides an effective method to control the anisotropy of
the atomic tunneling by laser intensities. The first Brillouin zone
of this system has also a hexagonal shape in the momentum space with
only two of the six corners in Fig.~1(d) are inequivalent,
corresponding to two different sites $A$ and $B$ in each cell in the
real hexagonal lattice, usually denoted as $K$ and $K'$. One can
choose ${\mathbf K}=(2\pi/a)(1/\sqrt{3},1)$ and ${\mathbf
K'}=-{\mathbf K}$, where $a=2\pi/(\sqrt{3}k_L)$ is the lattice
spacing. Taking a Fourier transform
$a^\dag_i=(1/\sqrt{N})\sum_{{\mathbf k}}{\rm exp}(i{\mathbf
k}\cdot{\mathbf A}_i)a_{\mathbf k}^\dag$ and
$b^\dag_j=(1/\sqrt{N})\sum_{{\mathbf k}}{\rm exp}(i{\mathbf
k}\cdot{\mathbf B}_j)a_{\mathbf k}^\dag$, where ${\mathbf A}_i$
(${\mathbf B}_j$) represents the position of the site in sublattice
$A$ ($B$) and $N$ is the number of sites of the sublattice, the
Hamiltonian (2) can be diagonalized and the eigenvalues have the
expression~[41]

\begin{widetext}
\vspace{0.2cm} $\displaystyle E_{\mathbf k}=\pm t
\sqrt{2+\beta^2+2\cos(k_y a)+4\beta \cos(\sqrt{3}k_x a/2)\cos(k_y
a/2)}.$\hfill (3) \vspace{0.2cm}
\end{widetext}

\vspace{0.25cm} As plotted in Fig. 1(e) and 1(f), there are two
branches of the dispersion relation, corresponding to the $\pm$ sign
in Eq. (3). When $0<\beta<2$, the two branches touch each other, and
around the touching points there appears a Dirac cone structure. One
has the same standard Dirac cones as the graphene material with
$\beta=1$ ~[35,63,64], and the cones squeeze in the $x$ or $y$
direction as $\beta$ deviates from $1$, but they still touch each
other. When $\beta>2$, a finite energy gap $\Delta_g=|t|(\beta-2)$
appears between the two branch. So, across the point $\beta=2$, the
topology of the Fermi surface changes, corresponding to a quantum
phase transition without any usual symmetry breaking~[65]. Such
topological phase transition associated with pair production
(annihilation) events has been investigated in Ref.~[66]. The
evolution of the Dirac points in the hexagonal lattice by varying
the asymmetry hopping and the resulting phase transition was also
studied in Ref.~[44]. With this phase transition, the system changes
its behavior from a semimetal to an insulator at the half filling
case (means one atom per cell; note that each cell has two sites).
Around the half filling, the Fermi surface is close to the touching
points, and one can expand the momentum ${\mathbf k}$ around one of
the touching points ${\mathbf K}\equiv(k_x^0,k_y^0)$ as
$(k_x^0,k_y^0)=(k_x^0+q_x,k_y^0+q_y)$. Up to the second order of
$q_x$ and $q_y$, the dispersion relation (3) becomes

\vspace{0.25cm}$E_{\mathbf q}=\pm \sqrt{\Delta_g^2+v_x^2 q_x^2+v_y^2
q_y^2}$,\hfill (4)

\vspace{0.25cm} \noindent where $\Delta_g=0$, $v_x=\sqrt{3}\beta t
a/2$, and $v_y=t a \sqrt{1-\beta^2/4}$ for $0<\beta<2$;
$\Delta_g=|t|(\beta-2)$, $v_x= t a \sqrt{3\beta/2}$, and $v_y=t a
\sqrt{\beta/2-1}$ for $\beta>2$. This simplified dispersion relation
$E_{\mathbf q}$ is actually a good approximation (named as long
wavelength approximation) as long as $q_x$, $q_y\lesssim 1/2a$.
Compared with the standard energy-momentum relation for the
relativistic Dirac particles, here $\Delta_g$ and $v_{x,y}$ take the
meaning of rest energy and the velocity of light respectively. The
wave function for the quasiparticles around the half filling then
satisfies the Dirac equation $i\hbar\partial_t \Psi={\mathcal H}_D
\Psi$, where the relativistic Hamiltonian ${\mathcal H}_D$ is given
by

\vspace{0.25cm}${\mathcal H}_D=v_x\sigma_x p_x+v_y\sigma_y
p_y+\Delta_g\sigma_z$,\hfill (5)

\vspace{0.25cm} \noindent where $\sigma_{x,y,z}$ are the three Pauli
matrices.

Through an analogy to the graphene physics, we have shown that by
controlling the lattice anisotropy, one can realize both massive and
massless Dirac fermions and observe the phase transition between
them. This proposal was recently proved to be experimentally
feasible in Ref.~[45], where the temperature requirement and
critical imperfections in the laser configuration are considered in
detail. Even in the presence of a harmonic confining potential, the
Dirac points are also found to survive~[67]. In the presence of
atomic interactions, the many-body physics of Dirac particles in
graphene-type lattices, such as novel BCS-BEC crossover~[42],
topological phase transition between gapless and gapped
superfluid~[47] and even charge and bond ordered states with
$p$-orbital band of lattices~[43,68], have been investigated.
Notably, the realization of ultracold quantum gases in a hexagonal
OL was reported in a very recent experiment~[69], which paves the
important way to mimic the relativistic Dirac fermions and the
aforementioned beyond-graphene physics with controllable systems.

\subsubsection{Pseudospin-1 massless Dirac fermions: Atoms in a
$\mathcal{T}_3$/line-centered-square OL}

\noindent It is interesting to note that cold atoms in OLs with
other structures of symmetries can also simulate the 2D relativistic
Dirac fermions. One of such examples was proposed in Ref.~[48]. It
showed that atoms trapped in the $\mathcal{T}_3$ lattice can behave
as the massless Dirac fermions with pseudospin $S=1$, instead of
$S=1/2$ for those in the hexagonal lattice. The so-call
$\mathcal{T}_3$ lattice, illustrated in Fig. 2(a), has a unit cell
with three different lattice sites, one sixfold coordinated site H
and two threefold coordinated sites A and B.

\vspace{0.8cm}

The $\mathcal{T}_3$ OL can be experimentally realized through three
counterpropagating pairs of laser beams with the same wavelength
$3/2a$ and linearly polarized with the electrical field in the
$x$-$y$ plane, which are similar with the laser setup for creating a
hexagonal OL. Given a polarization of a pair of lasers on the $y$
axis, the other two pairs are obtained by rotating $2\pi/3$ around
the z axis. The Schr\"{o}dinger equation for the $\mathcal{T}_3$ OL
filled with fermionic atoms in the tight-binding approximation is
given by

\vspace{0.15cm} \noindent $\displaystyle E\Psi_H({\mathbf
R}_H)=-t\sum_j \Psi_A({\mathbf R}_H+\tau_j)+\Psi_B({\mathbf
R}_H+\tau_{j+1}),$\hfill(6a)

\vspace{0.15cm}\noindent $\displaystyle
 E\Psi_{\alpha}({\mathbf R}_{\alpha})=-t\sum_j \Psi_H({\mathbf R}_{\alpha}-\tau_j),~~~~~\alpha \in \{A,B\}$.\hfill (6b)

\vspace{0.25cm} \noindent Here $\Psi_{\alpha}({\mathbf R}_{\alpha})$
is the amplitude of the wave function on sublattice $\alpha$
($=A,H,B$), and the $\tau_j$ connect nearest neighbors. Solving the
Eq. (6), one can obtain the energy-momentum relationship

\begin{widetext}
\vspace{0.2cm} $\displaystyle E_0({\mathbf k})=0,$\hfill(7a)

\vspace{0.15cm} $\displaystyle
 E_{\pm}({\mathbf k})=\pm t \sqrt{6+4\{\cos[({\mathbf v}_1-{\mathbf v}_2)\cdot{\mathbf k}]+\cos[{\mathbf v}_1 \cdot {\mathbf k}]+\cos[{\mathbf v}_2 \cdot
{\mathbf k\}]}}$~~, \vspace{0.3cm} \hfill (7b)
\end{widetext}

\vspace{0.25cm} \noindent where $E_{\pm}$ exhibit two Dirac points,
${\mathbf K}$ and ${\mathbf K'}$ as seen in Fig. 2(b), and $E_0$
represents a flat band. For the near half-filling case, one arranges
the component $\alpha=\{A,H,B\}$ into a pseudospin triplet and
expands the momentum ${\mathbf k}$ in the vicinity of ${\mathbf K}$,
and thus obtains the effective Hamiltonian of Dirac-Weyl form given
by

\vspace{0.25cm}${\mathcal H}_{DW}=v_F{\mathbf S}\cdot {\mathbf
p}$,\hfill (8)

\vspace{0.25cm} \noindent which describes the massless Dirac
fermions with total pseudospin $S=1$. Here $v_F=3ta/\sqrt{2}$ is the
Fermi velocity, ${\mathbf p}$ is the momentum operator in the $xy$
plane, and the pseudospin vector operator ${\mathbf
S}=(S_x,S_y,S_z)$ ($S_{x,y,z}$ are $3\times 3$ matrices which
satisfy angular momentum commutation relations) reflected the three
inequivalent lattice sites per unit cell in $\mathcal{T}_3$. A
comparison on the topological properties between the hexagonal
lattice and the $\mathcal{T}_3$ lattice was given in Ref.~[49],
where the former is shown to be topological insulator and the latter
to be trivial insulator in the framework of quantum spin Hall
states.

\vspace{5mm}
\includegraphics[width=7cm,height=3cm]{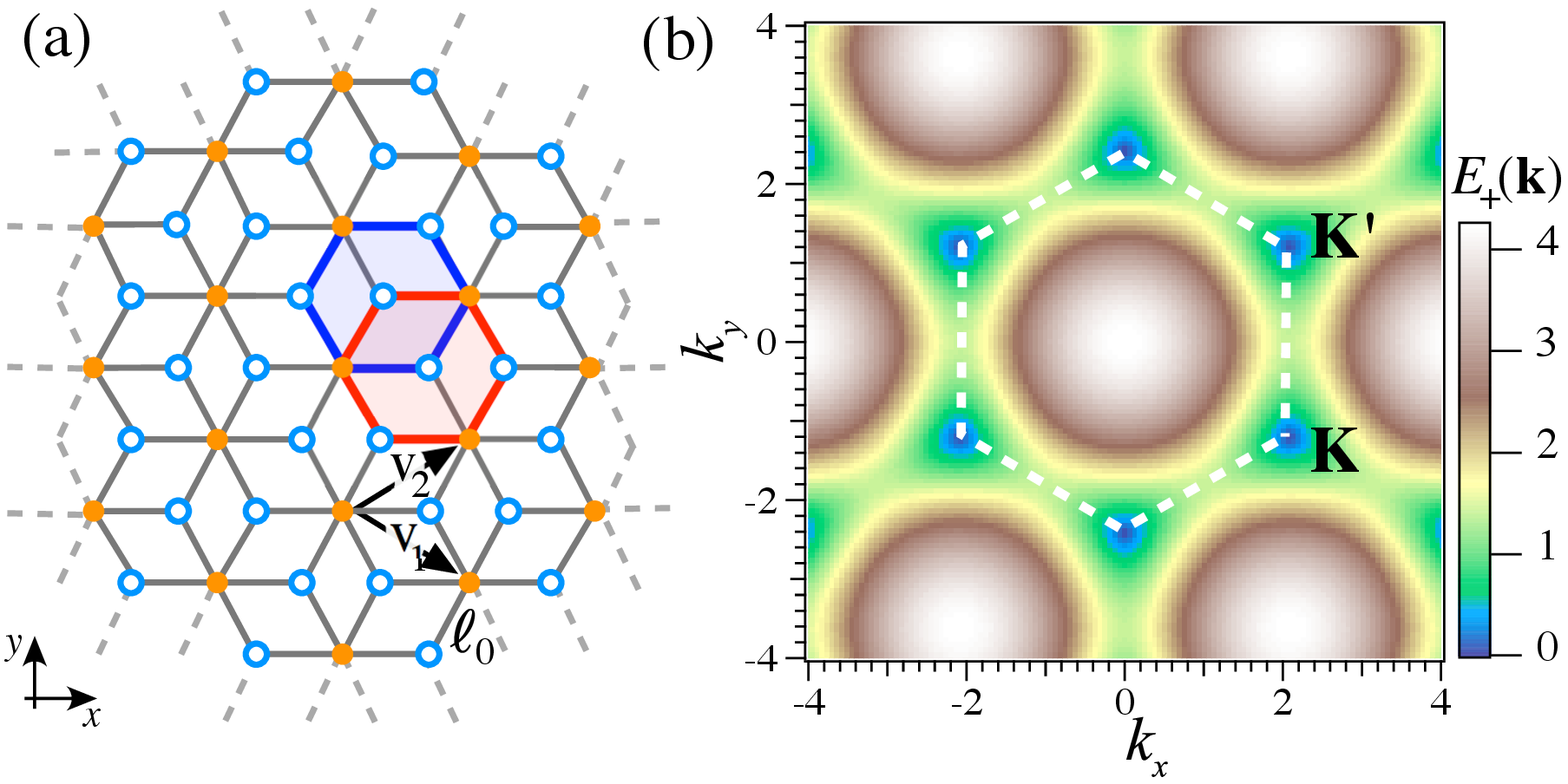} \vspace{1mm}
{\baselineskip
10.5pt\renewcommand{\baselinestretch}{1.05}\footnotesize

\noindent{\bf Fig.~2}\quad (a) The $\mathcal{T}_3$ lattice. It is
characterized by translation vectors ${\mathbf
v}_1=(3/2;-\sqrt{3}/2)a$ and ${\mathbf v}_2=(3/2;\sqrt{3}/2)a$. Open
circles mark the two sublattices $A$ and $B$, forming a hexagonal
lattice. Solid circles mark the hub sites $H$ forming a (larger)
triangular lattice. (b) Contour plot of $E_+({\mathbf k})$ in units
of the hopping energy $t$, cf. Eq. (7b). The dashed hexagon defines
the first Brillouin zone, ${\mathbf K}=2\pi a^{-1}(1/3;-\sqrt{3}/9)$
and ${\mathbf K'}=2\pi a^{-1}(1/3;\sqrt{3}/9)$ are two nonequivalent
Dirac points. Reproduced from Ref.~[48], Copyright \copyright~2009
the American Physical Society.

}\vspace{5mm}

\vspace{3mm}
\includegraphics[width=7cm,height=6cm]{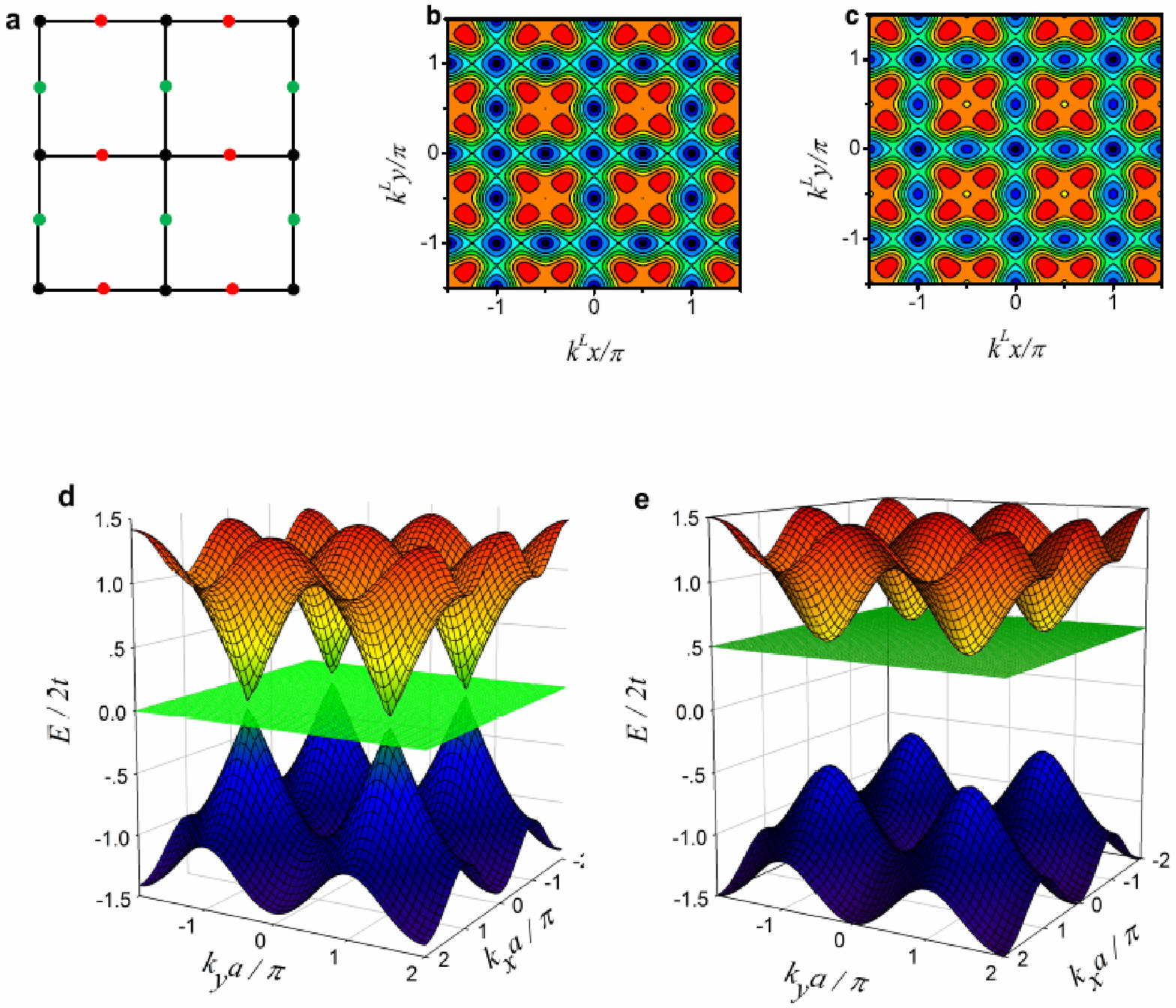} \vspace{1mm}
{\baselineskip
10.5pt\renewcommand{\baselinestretch}{1.05}\footnotesize

\noindent{\bf Fig.~3}\quad (a) Schematic illustration of the LCS
lattice, energy contour of the optical potential in Eq. (9) with (b)
$V_1=2V_2$ and (c) $V_1=2.2V_2$, and dispersion relation on the LCS
lattice with (d) $\Delta'_g=0$ and (e) $\Delta'_g=t$. Reproduced
from Ref.~[50], Copyright \copyright~2010 the American Physical
Society.

}\vspace{2mm}

Another example of quantum simulation of 2D massless Dirac fermion
was proposed to be realized in a LCS OL~[50]. The massless Dirac
fermions there also have pseudospin-1, and may exhibit a perfect
all-angle KT (we will discuss KT in the Section IV), i.e., the
barrier is completely transparent for all incident angles~[50]. The
LCS lattice is schematically illustrated in Fig. 3(a), with three
sublattices denoted by the black (sites $A$), red (sites $B$), and
green (sites $C$) points, respectively. The LCS lattice can be
realized in the ultracold atomic system by applying six detuned
standing-wave laser beams, four of which are applied along the ${\bf
e}_x$ and ${\bf e}_y$ directions with optical wave vectors $k_L$ and
$2k_L$, respectively; the other two are applied along the $({\bf
e}_x+{\bf e}_y)/\sqrt{2}$ directions with optical wave vector $2k_L$
and relative phase $\pi/2$, respectively. Thus the whole optical
potential is

\vspace{0.25cm}\noindent$\displaystyle V(x,y) = V_1[\sin^2(k_L
x)+\sin^2(k_L y)+\sin^2(2k_L x)$

 ~~~~~~~~~~$\displaystyle +\sin^2(2k_L y)]+V_2\{
\sin^2\left[k_L (x+y)+\frac{\pi}{2}\right]$

 ~~~~~~~~~~$\displaystyle \cdot\sin^2\left[k_L
(x-y)+\frac{\pi}{2}\right] \}$ \hfill (9)

\vspace{0.25cm} \noindent with tunable potential amplitudes $V_1$
and $V_2$, and lattice constant $a=\pi/k_L$. Two examples of energy
contours of LCS OL are plotted in Fig. 3(b) and 3(c), in which the
potential minima are marked with the black and blue points. For
$V_1=2V_2$, $V(x,y)$ is the same at sites $A$-$C$ so that their site
energies $\epsilon_A=\epsilon_B=\epsilon_C$. While for $V_1\neq
2V_2$, $\epsilon_B=\epsilon_C\neq\epsilon_A$ and one may set
$\epsilon_B=-\epsilon_A=\Delta'_g$ for the symmetry consideration.
The Hamiltonian of LCS lattice ${\mathcal H}_{LCS}=\sum_i \epsilon_i
c^{\dag}_i c_i+t\sum_{\langle i,j \rangle} c^{\dag}_i c_j$ can be
diagonalized with three branches through a similar procedure for Eq.
(2). One is a flat band with energy $E_0=\Delta'_g$, and the other
two dispersive bands are
$E_{\pm}=\pm\sqrt{\Delta'^2_g+4t^2[\cos^2(k_x a/2)+\cos^2(k_y
a/2)]}$. Typical energy bands are shown in Fig. 3(d) for
$\Delta'_g=0$ and Fig. 3(e) for $\Delta'_g=t$.

At $\Delta'_g=0$, only one nonequivalent Dirac point appears at
${\mathbf K}=(\pi/a,\pi/a)$. In the vicinity of the Dirac cone, the
ultracold atoms behave as the massless Dirac fermions with
pseudospin-1, described by Eq. (8) with $v_F=ta$ in this case.
Interestingly, the massless Dirac fermions in the HSC lattice are
topologically different from those in the hexagonal and
$\mathcal{T}_3$ OLs presented previously because there are not two
nonequivalent Dirac cones but a single one in the first Brillouin
zone. Note that the solid-state material with a single (or odd
number) Dirac cone, named as strong topological insulator, has
attract considerable attention recently~[37]. Besides, the flat band
with Dirac cones meeting at the same energy, which means the
particle can have an infinite effective mass (flat band) or a zero
effective mass (Dirac fermions), may result in interesting atom
dynamics in OLs, such as different behaviors of atomic localization
for bosons and fermions~[70].

\subsubsection{Dirac fermions in a square OL with a gauge field}

\noindent The massless Dirac fermions can also be simulated by using
ultracold atoms in a simple square OL subjected to certain
light-induced synthetic gauge fields (which will be addressed in
Section III). The synthetic fields are necessary because the energy
bands of a single square lattice do not exhibit the Dirac cone
structure.

\vspace{3mm}
\includegraphics[width=7cm,height=6.5cm]{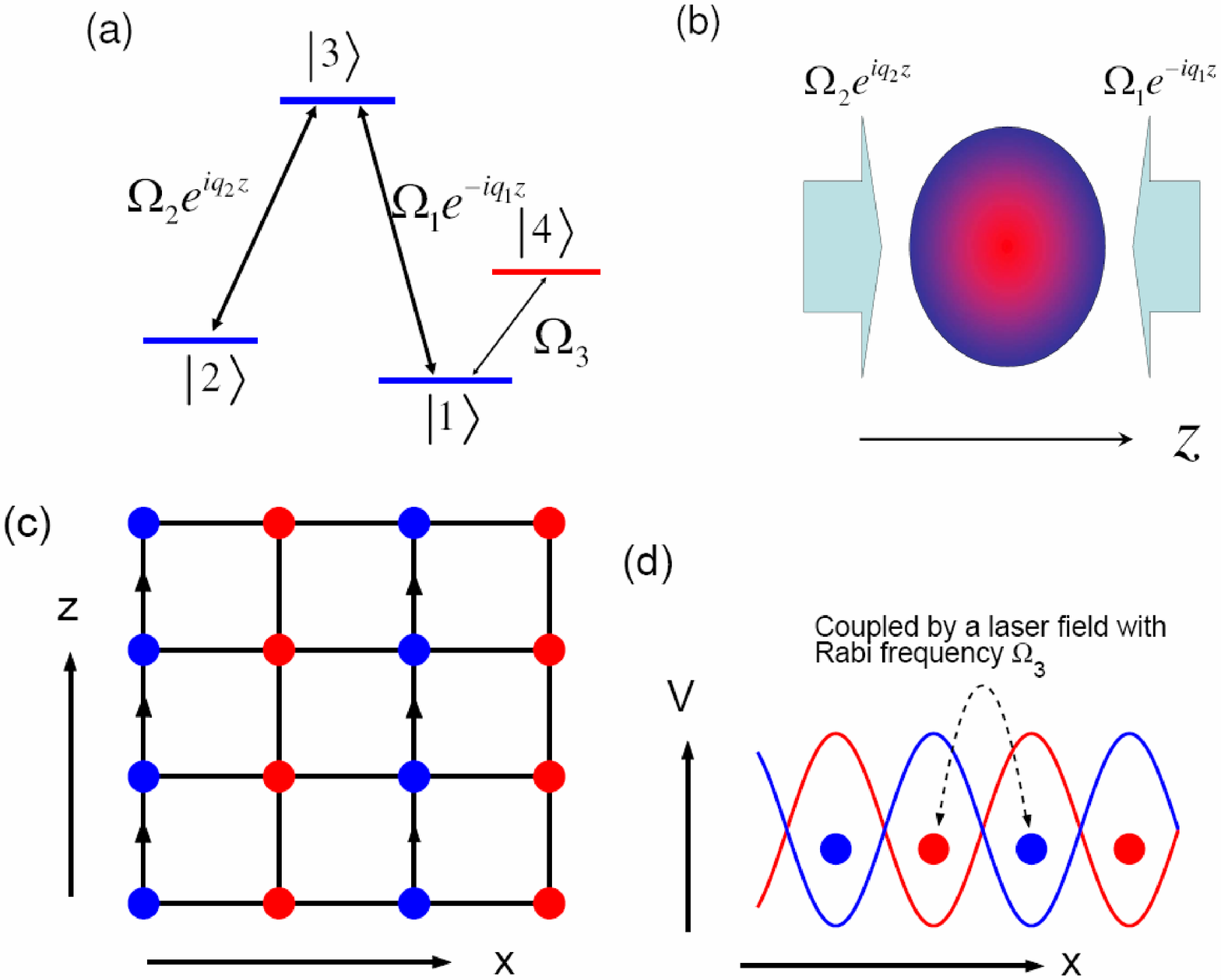} \vspace{1mm}
{\baselineskip
10.5pt\renewcommand{\baselinestretch}{1.05}\footnotesize

\noindent{\bf Fig.~4}\quad (a) The atomic levels and the
interactions between atoms and laser fields. (b) Schematic
representation of the experimental setup with the two laser beams
incident on the cloud of atoms. (c) Schematic of the square optical
lattice and the designed phase factor (denoted by arrows). (d) The
scheme of overlapping the two state-selective optical lattices.
Reproduced from Ref.~[52], Copyright \copyright~2009 the American
Physical Society.

}\vspace{2mm}

The first scheme along this line was proposed in Ref.~[52], where
the fermionic atoms are loaded in a square OL subjected to a U(1)
gauge field generated by laser-atom interactions. As shown in Fig.
4(a) and (b), the noninteracting atoms of mass $m$ with four levels
in the $x$-$z$ plane are coupled with three laser beams. The ground
state $|1\rangle$ is coupled to the excited state $|3\rangle$ via a
laser field with the corresponding Rabi frequency $\Omega_1
e^{-iq_1z}$ and the state $|2\rangle$ is coupled to the excited
state $|3\rangle$ via a laser field with the corresponding Rabi
frequency $\Omega_2 e^{iq_2z}$. The cold atoms are trapped in two
state-selective optical potentials as shown in Fig. 4(c) and (d),
say atoms with internal states $|1\rangle $, $|2\rangle$, and
$|3\rangle$ are trapped in odd columns (sublattice $A$), while atoms
with internal states $|4\rangle$ are trapped in even columns
(sublattice $B$). The two sublattices make up a 2D rectangular
lattice with the lattice spacings $a_x$ and $a_z$, especially a 2D
square lattice for $a_x=a_z$. In the basis $\{|1\rangle,
|2\rangle,|3\rangle,|4\rangle\}$, the total Hamiltonian of this
system can be written as

\vspace{0.25cm} \noindent ${\mathcal H}_{S}= {\hat H_0}+\hbar \left(
  \begin{array}{cccc}
    V_A & 0 & \Omega_1e^{iq_1z} & \Omega_3 \\
    0 & V_A & \Omega_2e^{-iq_2z} & 0 \\
    \Omega_1e^{-iq_1z} & \Omega_2e^{iq_2z} & V_A & 0 \\
    \Omega_3 & 0 & 0 & V_B \\
  \end{array}
\right)$,\hfill (10)

\vspace{0.25cm} \noindent where ${\hat
H_0}=-\frac{\hbar^2}{2m}\nabla^2$ is kinetic energy operator, and
$V_{A,B}$ are the two state-selective periodic potentials.
Diagonalizing the Hamiltonian (10) yields two degenerate dark states
(with eigenenergy $E=0$)

\vspace{0.25cm}$\displaystyle
 |\chi_1\rangle=\cos\theta| 1\rangle -\sin\theta e^{iqz}| 2\rangle,$\hfill
(11a)

\vspace{0.15cm}$\displaystyle |\chi_2\rangle=|4\rangle$,\hfill (11b)

\vspace{0.25cm} \noindent and two bright states, where $q=q_1+q_2$
and $\tan\theta=|\Omega_1|/|\Omega_2|$. If the motions of the atoms
that initially prepared in the dark state subspace satisfy the
adiabatic condition, one can safely use the adiabatic approximation
and reduce the Hamiltonian to

\vspace{0.25cm}$\displaystyle \hat H=\int d^2r
\hat\Phi_{1}^\dag\left[\frac{1}{2m}(-i\hbar\nabla-{\bf
{A}})^2+{V}_{A}\right]\hat\Phi_{1}$

\vspace{0.15cm}~~~~~~$\displaystyle +\int d^2r
\hat\Phi_{4}^\dag\left[-\frac{\hbar^2}{2m}\nabla^2+{V}_{B}\right]\hat\Phi_{4}$

\vspace{0.15cm}~~~~~~$\displaystyle +\hbar\Omega_e\int d^2r
\left(\hat\Phi_{4}^\dag\hat\Phi_{1}+\hat\Phi_{1}^\dag\hat\Phi_{4}\right)$,\hfill
(12)

\vspace{0.25cm} \noindent in the second quantized form, using
$\{|\chi_1\rangle,|\chi_2\rangle \}$ as the basis. Here
$\hat\Phi_{i}(\bf r)$ and $\hat\Phi_{i}^\dag(\bf r)$ are field
operators corresponding to annihilating and creating an atom with
the internal quantum state $|i\rangle$ at coordinate position $\bf
r$ respectively, $\Omega_e=\Omega_3 \cos\theta$ and the U(1)
adiabatic gauge potential ${\bf A}=\hbar q \sin^2\theta {\bf e}_z$.
Taking the tight-binding limit, one can superpose the Bloch states
to get Wannier functions $w_a(\bf r-\bf r_i)$ and $w_b(\bf r-\bf
r_j)$ for sublattice $A$ and $B$, respectively. Expanding the field
operator in the lowest band Wannier functions as

\vspace{0.25cm}$\displaystyle \hat\Phi_{1}({\bf
r})=\sum_{m(odd),n}\hat{a}_{m,n}e^{\frac{i}{\hbar}\int_0^{{\bf
r}_{mn}}{\bf A}\cdot d{{\bf r}}}w_a({\bf r}-{\bf r}_{mn})$,\hfill
(13a)

\vspace{0.15cm}$\displaystyle\hat\Phi_{4}({\bf
r})=\sum_{m(even),n}\hat{b}_{m,n}w_b({\bf r}-{\bf r}_{mn})$,\hfill
(13b)

\vspace{0.25cm} \noindent and then substituting them into Eq. (12),
one can rewrite the Hamiltonian as

\vspace{0.25cm}$\displaystyle\hat H=-t\sum_{[m(odd),n]} \hat
b^\dag_{m+1,n+1}\hat b_{m+1}+e^{i\gamma}\hat a^\dag_{m,n+1}\hat
a_{m,n}$

\vspace{0.15cm}~~~~~~~~~~~~~~~~~~~~~$\displaystyle + 2\hat
a^\dag_{m,n}\hat b_{m+1,n}+{\rm H.c.}$,\hfill (14)

\vspace{0.25cm} \noindent where $\gamma=2\pi \hbar a_z\sin^2\theta $
is the phase resulted from the adiabatic gauge potential. Here the
isotropic atomic tunneling is assumed, which can be realized by well
adjusting the intensities of laser beams. Diagonalizing the
Hamiltonian (14) with $\gamma = \pi$ and $a_x=a_z=a$, one obtains
the quasiparticle energy spectrum $ E({\bf k})= \pm
2t\sqrt{\cos^2(k_x a)+\cos^2(k_z a)}$, which exhibits two
inequivalent Dirac points in the first Brillouin zone with one at
${\bf K}=\left({\pi}/{2a},{\pi}/{2a}\right)$. In the vicinity of the
Dirac point ${\bf K}$, the atoms behave like the massless Dirac
fermions described by the effective Dirac-like Hamiltonian

\vspace{0.25cm}$\displaystyle \hat{\cal H}= \hbar
v_F(\hat{p}_x\sigma_x+\hat{p}_z\sigma_z)$,\hfill (15)

\vspace{0.25cm} \noindent where $v_F=2ta/\hbar$ is the Fermi
velocity of this system.

The 2D massless Dirac fermions also emerge when ultracold fermionic
atoms are trapped in a square OL subjected to a U(2) non-Abelian
gauge field~[56], that is to say, the phase factor $\gamma$ in Eq.
(14) is replaced by a $2\times 2$ matrix. This kind of non-Abelian
OLs could be created by laser-assisted tunneling for atoms in
optical superlattices~[71-73]. They are characterized by a constant
Wilson loops, and the single-particle spectrum may depict a complex
structure termed as the Hofstadter "butterfly"~[74]. In the
proposal, the half-integer quantum Hall effect that has been
observed in the graphene [35,36] and the squeezed Landau vacuum due
to anisotropic of external gauge filed are also investigated.

The external gauge filed can even be replaced by a time-dependent
optical potential~[53,54], which acts to shake the lattice, leading
to the modification of effective tunneling strength. Such driven
tunneling can be used to generate an artificial staggered magnetic
field for atoms in a two-dimensional square optical lattice. For
bosonic gas in this optical system, the zero-temperature phase
diagram was shown to exhibit a finite-momentum superfluid phase with
a quantized staggered rotational flux~[53]. For fermionic gas, 2D
massless Dirac fermions may also be provided~[54]; and mixing both
bosonic and fermionic atoms, one may even reach the strongly
interacting regime for 2D Dirac fermions~[55], which allows to
investigate the transition between the Dirac and the Fermi liquid,
and some correspondences with high-temperature cuprates.

Another model with a different time-dependent optical potential,
which is also used to create an artificial magnetic field in cold
atoms loaded on a square OL, was recently presented in Ref.~[58].
The effective dynamics of low energy quasiparticles in that system
can be described by a Dirac Hamiltonian for massless fermions with
the unusual property that chiral symmetry is broken in the tunneling
energy term, leading to splitting the doubly degenerate Dirac cones
into two with tunable slopes. The two slopes describe two speeds of
light for the relativistic quasiparticles.

\vspace{3mm}
\includegraphics[width=3.5cm,height=3.5cm]{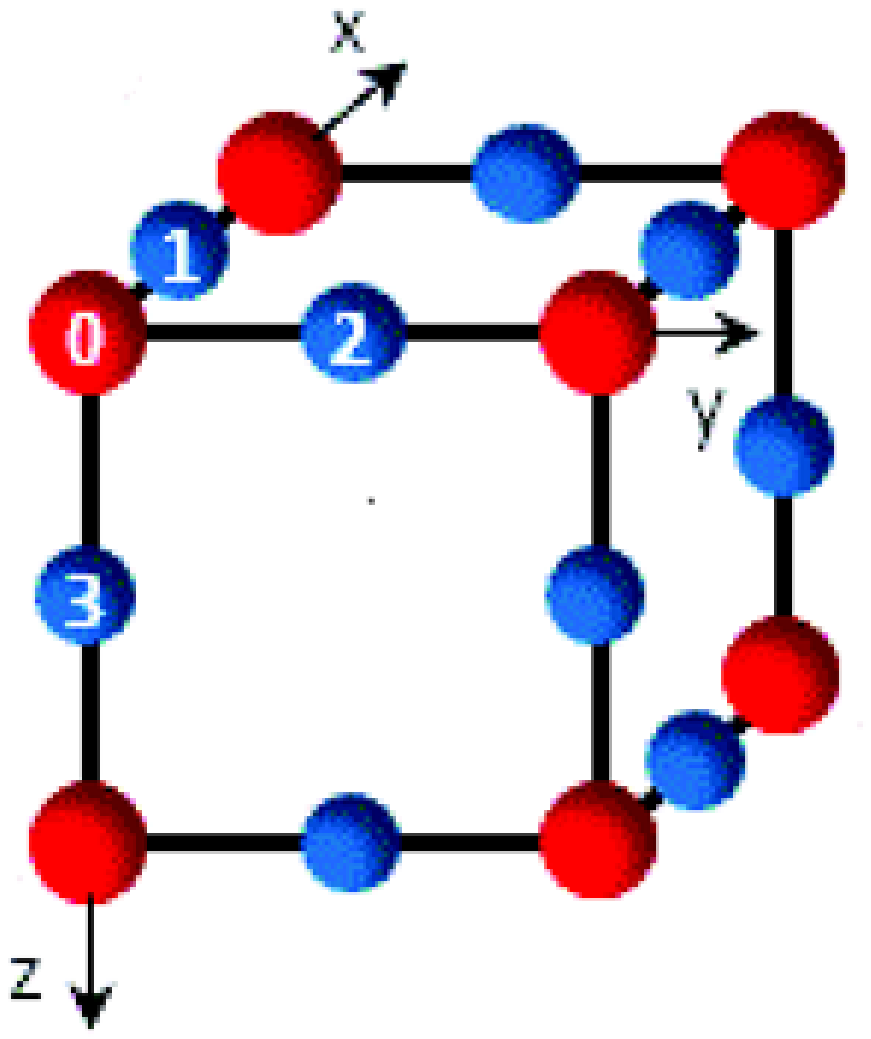} \vspace{1mm}
{\baselineskip
10.5pt\renewcommand{\baselinestretch}{1.05}\footnotesize

\noindent{\bf Fig.~5}\quad A crystal cell of the edge-centered cubic
lattice and the axis system adapted in this paper. One crystal cell
includes four inequivalent lattice sites labeled by 0, 1, 2 and 3,
respectively. Reproduced from Ref.~[59], Copyright \copyright~2010
the American Physical Society.

}\vspace{2mm}

\subsection{Simulation of 3D relativistic Dirac fermions}

\noindent The above mentioned work are all limited in 2D Dirac
fermions, and the 3D ones in the OLs would be valuable to explore
since the 3D Dirac fermions differ from 2D ones by transport
properties, localization, etc. We review in the following that cold
atoms loaded in an edge-centered cubic (ECC) OL can be used to
simulate the 3D Dirac-like fermions~[59].

The ECC lattice as shown in Fig. 5 is the three dimensional
counterpart of the LCS. In certain conditions (established by Eq.
(19) below), the cold atoms on the lattice sites can be regarded
approximately as the ground states of anisotropic 3D harmonic
oscillators with energy
$\epsilon^{(i)}_g=\frac12\hbar\sum_{\nu=x,y,z}\omega^{i}_\nu$, where
$\omega_\nu^{(i)}=[\partial^2_\nu V/m]^{1/2}_{\rm i}$ is the
frequency of the oscillator on site $i$ along the axis $\nu$ with
potential $V$(see Ref.~[59] for its expression). Substituting $V$
into the expression of $\omega^{i}$, yields

\vspace{0.25cm}$\displaystyle \epsilon_g^{(0)}= E_r\left[\frac32
\sqrt{\alpha_1-4\alpha_2+4\alpha_3} \right]$,\hfill (16a)

\vspace{0.15cm}$\displaystyle
\epsilon_g^{(1)}=E_r\left[\sqrt{\alpha_1+4\alpha_3} + \frac12
\sqrt{-\alpha_1+4\alpha_2+4\alpha_3}\right]$,\hfill (16b)

\vspace{0.15cm}$\displaystyle \epsilon_g^{(3)}=
\epsilon_g^{(2)}=\epsilon_g^{(1)}$,\hfill (16c)

\vspace{0.25cm} \noindent where $\alpha_i=A_i/E_r (i=1,2,3)$ are
dimensionless parameters with $E_r=\hbar^2 (2\pi/a)^2/2m$ being the
recoil energy. The 3D harmonic potentials on sites 0 and 1 are
lifted by the bottom energies $V(0,0,0) = 6A_2$ and $V(a/2,0,0) =
A_1+2A_2$ on site 0 and site 1, respectively. The ground state
energies of these sites are

\vspace{0.25cm}$\displaystyle E_g^{(0)} = \epsilon_g^{(0)} +
6\alpha_2E_r$,\hfill (17a)

\vspace{0.15cm}$\displaystyle E_g^{(1)} = \epsilon_g^{(1)} +
(\alpha_1+2\alpha_2)E_r$.\hfill (17b)

\vspace{0.25cm} \noindent Since the ground state energies on lattice
sites $1$, $2$ and $3$ are the same, we may address only one of
them, such as $E_g^{(1)}$ in Eq. (17b). We limit ourself to the
energy band formed by the ground states of the lattice sites. To
this end, one has to tune the parameters $A_1,A_2$ and $A_3$ (or
$\alpha_1, \alpha_2$ and $\alpha_3$) carefully to satisfy the
following conditions

\vspace{0.25cm}$\displaystyle |E^{(0)}_g-E^{(1)}_g| \ll 2\min
[\epsilon^{(0)}_g,\epsilon^{(1)}_g]$,\hfill (18a)

\vspace{0.15cm}$\displaystyle |\epsilon^{(0)}_g+\epsilon^{(1)}_g|
\ll 2V_j$.\hfill (18b)

\vspace{0.25cm} \noindent The parameter $V_j$ is the well depth that
can be estimated as $V_j=[2V(0,0,a/4)-V(0,0,0)-V(0,0,a/2)]/2=A_3$.
Equation 19(a) ensures that other energy levels of the oscillators
on each sites are separated sufficiently far away and then the
energy bands are formed by the ground states, and Eq. 18(b)
guarantees that the calculated ground levels on all sites are much
lower than the well depth so as to local levels can be formed and
the ground energies can be evaluated by Eq. (16a)-(17b). One can
estimate that the parameter region with $V_1/V_2\approx4$ and
$V_3\gtrsim30E_r$ fulfills the above condition.

Under the above harmonic approximate conditions, the tight-binding
Hamiltonian of the system in momentum space can be worked out as

\vspace{0.25cm} $\displaystyle H_k =2t\left(
                     \begin{array}{cccc}
                       -\delta & \cos\Gamma_x & \cos\Gamma_y & \cos\Gamma_z \\
                       \cos\Gamma_x & \delta & 0 & 0 \\
                       \cos\Gamma_y & 0 & \delta & 0 \\
                       \cos\Gamma_z & 0 & 0 & \delta \\
                     \end{array}
                   \right)
$,\hfill (19)

\vspace{0.25cm} \noindent in the basis
$\{|0\rangle,|1\rangle,|2\rangle,|3\rangle\}$ with $|i\rangle$ being
the single particle ground state on site $i$. Here the mid-point of
the energy on sites $0$ and $1$ is chosen as the energy reference,
$\Gamma_{x,y,z}=k_{x,y,z} a/2$ with wave vector $k_i$ ($i=x,y,z$)
along the direction $i$, $2\delta=E^{(1)}_g-E^{(0)}_g$ is the energy
difference between site 1 and site 0, and $t$ is the hopping
constant. Diagonalizing $H_k$, one obtains the four band dispersions
given by

\vspace{0.25cm} $\displaystyle E_1=E_2 = \delta$, \hfill (20a)

\vspace{0.15cm} $E_{\pm}=\pm \sqrt{ 4t^2(\cos^2\Gamma_x + \cos^2
\Gamma_y + \cos^2\Gamma_z) + \delta^2 }$. \hfill (20b)

\vspace{0.25cm} \noindent  The first two dispersions are flat bands
stick to the bottom of the upper band or the top of the lower band,
depending on the sign of $\delta$. Note that 3D bulk flat band has
not attracted much attention, in contract to previous work based on
1D or 2D model~[50,56,70]. The last two dispersions possess the
characteristic of Dirac particles near the only one Dirac point
${\bf K}=(\pi/a,\pi/a,\pi/a)$ at the corner of the first Brillouin
zone, and they touch each other when $\delta=0$. At the vicinity of
the Dirac point, the dispersion (21b) is approximately rewritten as

\vspace{0.25cm} $\displaystyle  \tilde{E}_{\pm}=\pm
\sqrt{m^{*2}c^{*4}+p^2c^{*2}}$, \hfill (21)

\vspace{0.25cm} \noindent where $p=\hbar k$ is the momentum with
$k=(k_x^2+k_y^2+k_z^2)^{1/2}$ being the amplitude of 3D wave-vector,
$c^*=at/\hbar$ is the effective light speed, and
$m^*=\hbar^2\delta/(at)^2$ is the effective rest mass. The
dispersion describes 3D massive Dirac fermions, and reduced to
massless Dirac dispersion, i.e., $E_{\pm}=\pm c^* p$, when
$\delta=0$.

Apart from this scheme, it has been shown~[60] that the 3D
relativistic fermions can also be simulated with ultracold fermionic
atoms in a 3D optical superlattice as an extension system of 2D
non-Abelian OL in~[56]. Interestingly, the 3D relativistic fermions
simulated in this way are the so-call naive Dirac fermions in
lattice gauge theories~[75,76], and thus provide a realization of
the fermion doubling problem: Two massless Dirac fermions appear in
the Brillouin zone, each of them has a different chirality,
corresponding to the right and left-handed particles. Furthermore,
by tuning the laser intensities, the system may allow to decouple
the doublers from a single Dirac fermion through inverting their
effective mass, i.e., a quantum simulation of Wilson fermions~[77].
In this regime, the atomic gas corresponds to a phase of matter, 3D
topological insulators, where Maxwell electrodynamics is replaced by
axion electrodynamics~[37,78]. The chirality symmetry breaking via
mass terms is also crucial for simulating the Wess-Zumino
supersymmetry model~[34].

The 3D Dirac fermions using ultracold atoms are also proposed to be
realized in a cubic optical lattice subjected to a synthetic
frustrating magnetic field $\vec{B}=\pi \phi_0 (1,1,1)$ with
$\phi_0=\hbar/2ma^2$~[61]. The tight-binding Hamiltonian of the
system can be described by

\vspace{0.25cm} $\displaystyle  H = - t \sum_{\langle i, j \rangle}
\left( c_{i}^{\dag} e^{-i A_{ij} } c_{j} + h.c. \right) $,\hfill
(22)

\vspace{0.25cm} \noindent where the phase factor is
$A_{ij}=\frac{m}{\hbar}\int_i^j \vec{A}\cdot d\vec{l}$, with the
vector $\vec{A}=\pi(0,x-y,y-x)$ since $\vec{B}={\rm rot}\vec{A}$.
Diagonalization of Hamiltonian (22) yields the dispersion as

\vspace{0.25cm} $\displaystyle E_{\pm}=\pm 2t\sqrt{ (\cos^2 k_x a+
\cos^2 k_y a+ \cos^2k_z a)}$,\hfill (23)

\vspace{0.25cm} \noindent which exhibit two inequivalent Dirac
points again. Around the Dirac points, the atoms behave like the 3D
massless Dirac fermions. It is also shown to be possible to give
mass to the Dirac fermions by coupling the ultracold atoms to a
Bragg pulse, and to obtain a dimensional crossover from 3D to 2D
Dirac fermions by varying the anisotropy of the lattice.

\vspace{3mm}
\includegraphics[width=8cm,height=5cm]{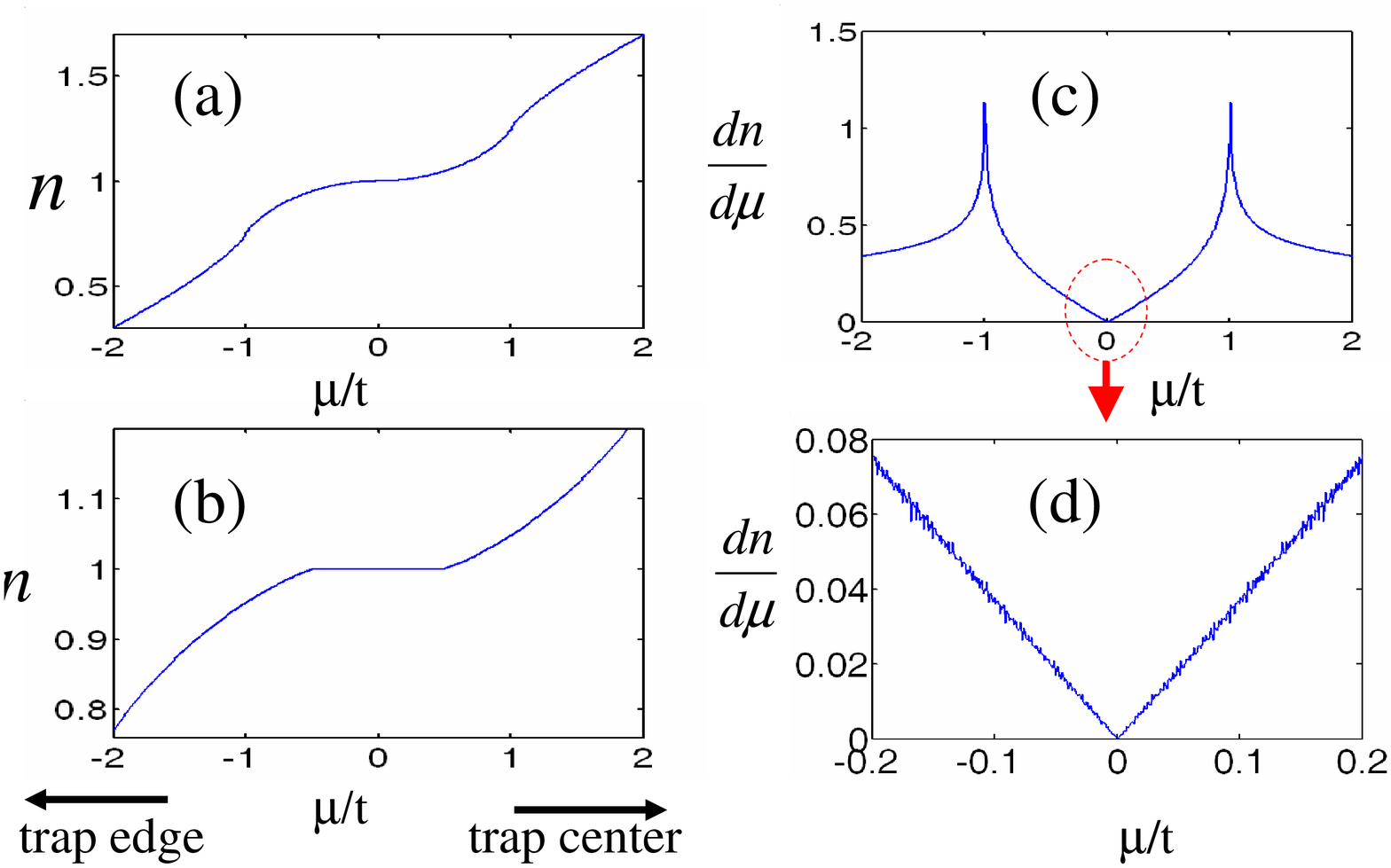}\vspace{1mm} {\baselineskip
10.5pt\renewcommand{\baselinestretch}{1.05}\footnotesize

\noindent{\bf Fig.~6}\quad The number density of atoms $n$ per unit
cell of the hexagonal lattice as a function of the chemical
potential $\protect\mu$ (corresponding to a re-scaled atomic density
profile in a trap) for (a) $\protect\beta =1$, and (b)
$\protect\beta =2.5$. A plateau with a width $2\protect\beta -4$
appears for the latter case which corresponds to the case when the
chemical potential sweeps inside the energy gap. (c) The derivative
$dn/d\protect\mu $ as a function of the chemical
potential $\protect\mu $ for $\protect\beta =1$. (d) An enlarged part of $%
dn/d\protect\mu $ at the vicinity of $\protect\mu =0$. The linearity
of the curve shows the linear dispersion relation for the
quasiparticles. Reproduced from Ref.~[41], Copyright \copyright~2007
the American Physical Society.

}\vspace{2mm}

\vspace{3mm}
\includegraphics[width=0.6\columnwidth]{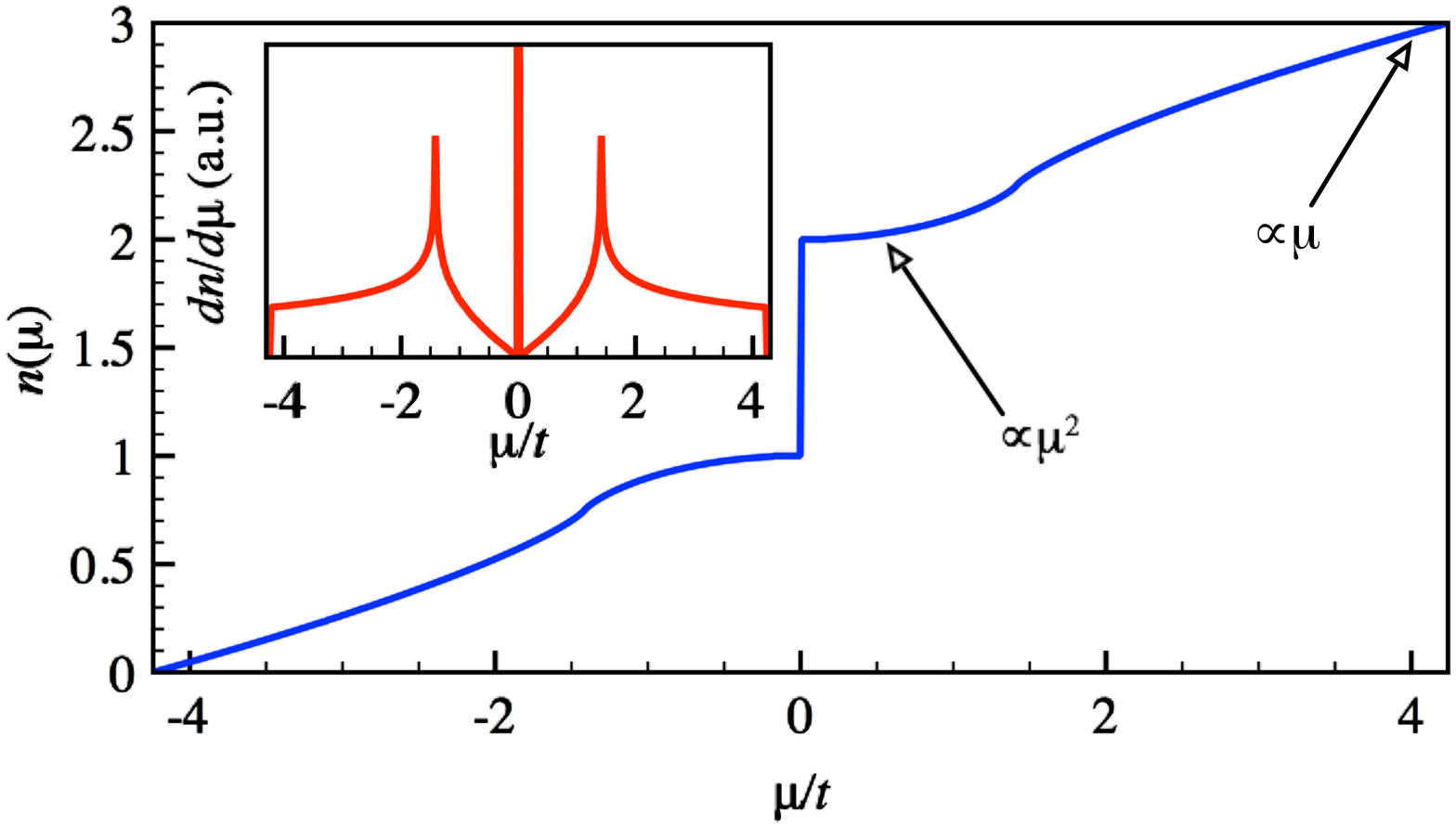}\vspace{1mm} {\baselineskip
10.5pt\renewcommand{\baselinestretch}{1.05}\footnotesize

\noindent{\bf Fig.~7}\quad The number of atoms $n$ per unit cell in
${\mathcal T}_3$ OLs at zero temperature as a function of $\mu/t$.
Inset: density of states, $dn/d\mu$, versus $\mu/t$. Reproduced from
Ref.~[48], Copyright \copyright~2009 the American Physical Society.

}\vspace{2mm}

\subsection{Detection of the Dirac quasiparticles}

\noindent We have shown that cold atoms trapped in OLs with a wide
range of structures can be used to simulate the Dirac equation, and
the quasiparticles in these systems behave as the relativistic ones.
Another important issue, however, is to experimentally verify the
existence of these relativistic quasiparticles. The widely used
detection technique based on the transport measurements for the
condensed-matter materials is typically not available for the atoms,
and nevertheless, there are some specific detection methods for the
trapped atomic gas. For simplicity, we focus on the aforementioned
hexagonal OL system~[41], and in the following we review two
different methods to confirm the relativistic quasiparticles and the
phase transition between massless and massive ones~[41]: the density
profile measurement~[79] and the Bragg spectroscopy~[80]. However,
the detection methods can be extended to other systems without loss
of generality.

{\sl Density profile measurement.---}The density profile of the
trapped atoms can be measured through the time-of-flight imaging
with the light absorption~[79]. Free fermions expand with ballistic
motion and from the final measured absorption images, one can
reconstruct the initial real-space density profile of the trapped
gas. Under the local density approximation, the local chemical
potential varies with the radial coordinate by $\mu =\mu
_{0}-V(\mathbf{r}),$ where $\mu _{0}$ is the chemical potential at
the trap center and $V(\mathbf{r})=m\omega ^{2}\mathbf{r}^{2}/2$ is
the global harmonic trapping potential. At temperature $T$, the
atomic density is given by

\vspace{0.25cm} $\displaystyle n(\mu )=\frac{1}{S_{0}}\int
f(k_{x},k_{y},\mu )dk_{x}dk_{y}, $ \hfill (24)

\vspace{0.25cm} \noindent where $S_{0}=8\pi ^{2}/\sqrt{3}a^{2}$ is
the area of the first Brillouin zone of the hexagonal lattice, and
$f(k_{x},k_{y},\mu )=1/\{\exp [(E_{\mathbf{k}}-\mu )/T]+1\}$
($E_{\mathbf k}$ in Eq.(3)) is the Fermi-Dirac distribution. At low
temperature ($T\sim 0)$, this density profile $n(\mu)$ is shown in
Fig. 6(a) and 6(b) for the parameters with massless and massive
Dirac quasiparticles, respectively. Clearly, a plateau at the atom
density $n=1$ in the density profile appears for the gapped phase
with massive Dirac fermions; but there is no such a plateau for the
case with massless Dirac fermions. So the plateau is associated with
massive quasiparticles, and its emergence provides an unambiguous
signal for the quantum phase transition between the two cases.
Furthermore, the linear dispersion relation for the massless Dirac
fermions can be confirmed by the derivative $dn/d\protect\mu $ since
one has
$\frac{%
\partial n}{\partial \mu }=\frac{4\pi }{v^{2}S_{0}}\left| \delta \mu \right|
$ around the Dirac cone for $\beta=1$ and $v_{x}=v_{y}=v$. As shown
in Fig. 6(c) and (d), $\frac{\partial n}{\partial \mu }$ is linearly
proportional to $\delta \mu $ indeed at the vicinity of the half
filling. So experimentally, from the measured density profile $n(\mu
)$, one can determine its slope, which signals the linear dispersion
relation, to confirm the massless Dirac fermions.

This method is also available for the detection of massless Dirac
fermion in ${\mathcal T}_3$ OLs~[48]. The similar results are shown
in Fig. 7, but with a sharp jump in the atomic density at $\mu=0$
due to the contribution from the highly degenerate flat band.

{\sl Bragg spectroscopy.---}The Bragg spectroscopy can provide an
alternative and complementary method to confirm the linear
dispersion relation for the massless Dirac fermions and the energy
gap for the massive ones~[41]. In Bragg spectroscopy~[80], one
shines two laser beams on the atomic gas as shown in Fig. 8(a). By
fixing the angle between the two beams (thus fixing the relative momentum transfer $\mathbf{%
q=k}_{2}-\mathbf{k}_{1}$, where $\mathbf{k}_{i}$ denotes the wave
vector of each laser beam), one can measure the atomic (or photonic)
transition rate by scanning the laser frequency difference $\omega
=\omega _{2}-\omega _{1}$. From the Fermi's golden rule, this
transition rate basically measures the following dynamical structure
factor~[80]

\vspace{0.25cm} $\displaystyle
S(\mathbf{q},\omega )=\sum_{\mathbf{k}_{1},\mathbf{k}_{2}}|\langle f_{%
\mathbf{k}_{2}}|H_{B}|i_{\mathbf{k}_{1}}\rangle |^{2}\delta \lbrack
\hbar \omega -E_{f\mathbf{k}_{2}}+E_{i\mathbf{k}_{1}}]$, \hfill (25)

\vspace{0.25cm} \noindent where $H_{B}=\sum_{\mathbf{k}_{1},\mathbf{k}_{2}}\Omega e^{i\mathbf{q\cdot r}%
}|i_{\mathbf{k}_{1}}\rangle \langle f_{\mathbf{k}_{2}}|+h.c.$
is the light-atom interaction Hamiltonian, and $|i_{\mathbf{k}%
_{1}}\rangle $ and $|f_{\mathbf{k}_{2}}\rangle $ denote the initial
and the final atomic states with the energies $E_{i\mathbf{k}_{1}}$
and $E_{f\mathbf{k}_{2}}$ and the momenta $\mathbf{k}_{1}$ and
$\mathbf{k}_{2}$, respectively. At the half filling, the excitations
are dominantly around the touching point, and we can use the
approximate dispersion relation in Eq. (4). For the isotropic case
($\beta =1$) with massless Dirac Fermions, $S(q,\omega)$ has the
expression

\vspace{0.25cm} $ S(q,\omega)=\left\{
\begin{array}{ll}
0, & (\omega \leq \omega _{r}) \\
\frac{\pi \Omega ^{2}}{8v_{F}}\frac{2q_{r}^{2}-q^{2}}{\sqrt{q_{r}^{2}-q^{2}}}%
, & (\omega >\omega _{r})
\end{array}
\right.$ \hfill (26)

\vspace{0.25cm} \noindent where $\omega _{r}=qv_F/\hbar $ ($q\equiv \left| \mathbf{q}\right| $) and $%
q_{r}=\hbar \omega /v_F$. This dynamical structure factor is shown
in Fig. 8(b). Note that in this case, the lower cutoff frequency
$\omega _{r}$ is linearly proportional to the momentum difference
$q$, and $\omega _{r}$ vanishes when $q$ tends to zero. The ratio
between $\omega _{r}$ and $q$ gives the Fermi velocity $v_F$, an
important parameter as the analogy of the light velocity for
conventional relativistic particles. For the anisotropic case with
$\beta >2$, the spectrum in Eq. (4) becomes quadratic with $E\approx
\pm (\Delta+\hbar ^{2}q_{x}^{2}/2m_{x}+\hbar ^{2}q_{y}^{2}/2m_{y})$
for small momentum transfer $\mathbf{q}$, where the effective mass
$m_{x,y}=\hbar ^{2}\Delta/v_{_{x},_{y}}^{2}$. The dynamical
structure factor in this non-relativistic limit becomes

\vspace{0.25cm} $ S(q,\omega )=\left\{
\begin{array}{ll}
0, & (\omega \leq \omega _{c}^{x,y}) \\
\frac{\pi \Omega ^{2}\Delta }{2v_{_{x}}v_{_{y}}}, & (\omega >\omega
_{c}^{x,y})
\end{array}
\right.$ \hfill (27)

\vspace{0.25cm} \noindent where $\omega _{c}^{x,y}=2\Delta +\hbar
^{2}q_{x,y}^{2}/4m_{x,y}$. Its form is shown in Fig. 8(b). The lower
cutoff frequency $\omega _{c}^{x,y}$ in this case does not vanish as
the momentum transfer goes to zero. This distinctive difference
between the dynamical structure factors in Eqs. (26) and (27) can be
used to distinguish the cases with massive or massless Dirac
fermions. From the variation of the cutoff frequency $\omega
_{c}^{x,y}$ as a function of the momentum transfer $q_{x,y}$, one
can also experimentally figure out the important parameters such as
the energy gap $\Delta $ and the effective masses $m_{x}$ and
$m_{y}$.

\vspace{3mm}
\includegraphics[width=7cm,height=4cm]{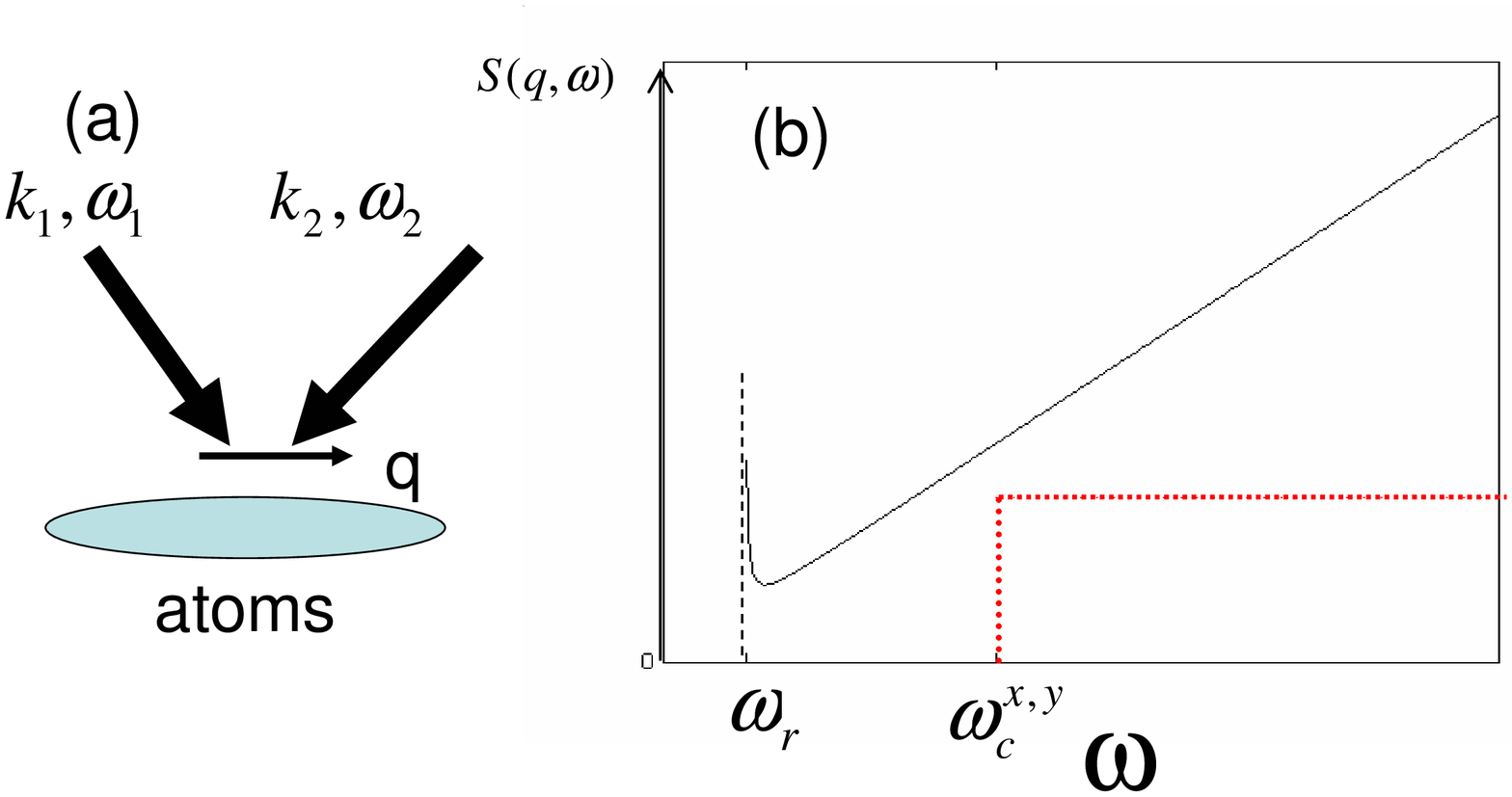} \vspace{1mm}
{\baselineskip
10.5pt\renewcommand{\baselinestretch}{1.05}\footnotesize

\noindent{\bf Fig.~8}\quad Schematic of the Bragg scattering. (a)
Atoms are illuminated by two laser beams with wave vectors $\mathbf{k}_{1}$, $\mathbf{k%
}_{2}$ and frequencies $\protect\omega _{1}$, $\protect\omega _{2}$,
respectively. (b) The dynamic structure factors $S(q,\protect\omega
)$ (not scaled) for the massless Dirac fermions (solid line) and for
the massive ones in the non-relativistic limit (dotted line). The
experimentally measurable quantities $\protect%
\omega _{r}$ and $\protect\omega _{c}^{x,y}$ give important
parameters for the quasiparticles. Reproduced from Ref.~[41],
Copyright \copyright~2007 the American Physical Society.

}\vspace{2mm}

\subsection{Quantum anomalous Hall effects}

\noindent As a unique electronic behavior of single layer graphene,
the relativistic Dirac fermions carrying charge in this 2D material
may result in novel transport phenomena. One of the most exotic
examples is that, when subjected to a perpendicular magnetic field,
the so-call half-integer quantum Hall effect (QHE) as a result of
the unconventional relativistic Landau level can be observed in
graphene~[35,36]. For cold atomic systems, such half-integer QHE is
principally able to be simulated since the relativistic particles
emerge~[56,61], and the additional requirement is an effective
uniform magnetic field, which can be generated by rotation or
optical dressing (will be discussed in Section III.A).

In contrast to the conventional QHE and the half-integer QHE, there
is another kind of QHE, called as quantum anomalous Hall effect
(QAHE) because the net magnetic flux is zero in each unit cell and
there are no Landau levels. However, all the three need the breaking
of time-reversal symmetry. Twenty years ago, Haldane proposed a toy
model to illustrate such QAHE as a result of the parity anomaly of
the 2D Dirac fermions in a hexagonal lattice~[81]. Nevertheless, it
is extremely hard to realize the Haldane's model experimentally in
ordinary condensed matter systems including the graphene because of
the unusual staggered magnetic flux assumed in the model. Other
schemes have been proposed to realize QAHE in semiconductor systems
with topological band structures~[82-84], where the time-reversal
symmetry is broken by the spin polarization. Besides, it is possible
to realize QAHE in graphene in the presence of both Rashba
spin-orbit coupling and an exchange field~[85], or in a
nanopatterned 2D electron gas~[86], or in a thin film of a
three-dimensional topological insulator with an exchange field~[87].
The QAHE states are promising for future device applications because
of its potential realization of dissipationless charge transport.

Interestingly, several proposals have been suggested to realize and
detect the QAHE by using ultracold atoms. An ingenious scheme is
proposed to realize the Haldane' model by using cold atoms trapped
in a hexagonal OL with an artificial staggered magnetic field (Berry
curvature) with hexagonal symmetry in Ref.~[88]. The wanted external
field can be achieved by coupling atomic internal state with the
standing waves of laser beams. While it remains a challenge to
experimentally realize such a staggered magnetic field. In
Ref.~[89], it was showed that the QAHE can be simulated with the
$p$-orbital band in the hexagonal OL through orbital angular
momentum polarization by applying a technique of rotating each
lattice site around its own center~[90]. An expanded version of this
proposal can be found in Ref.~[91]. Besides, it was recently
demonstrated to realize the QAHE with cold atoms trapped in an
anisotropic square OL which can be generated from available
experimental setups of double-well lattices with minor
modifications~[92]. The detection of the QAHE in cold atomic systems
can be achieved by direct detection of the Chern number with the
typical density-profile-measurement technique~[88], or alternatively
determining the topological phase transition from usual insulating
phase to quantum anomalous Hall phase by light Bragg scattering of
the edge and bulk states~[92]. All of these proposals involve
experimental techniques of implementing cold atoms to be developed.

\section{Simulation of Dirac equation with ultracold atoms in synthetic non-Abelian gauge fileds}

\noindent In the models discussed in Sec. II, the emergency of
relativistic Dirac fermions (equation) in OL systems generally
requires the long wavelength approximation, which supports the
continuous limitation. In this section, we introduce another kind of
proposals to simulate the tunable Dirac equation that are based on
generating effective gauge fields~[93-101] on bulk atomic gases,
which do not need OLs and thus operates in a continuous
regime~[102-105]. We will see that the effective relativistic
dynamics emerge again in the cold atom systems by using atom-light
interaction to generate effective gauge potentials acting on these
neutral atoms. Besides, there is no filling requirement in these
continuous systems, and thus the relativistic Dirac dynamics are
equivalent for both noninteracting bosonic and fermionic in the
single-particle physics. Moreover, it even holds for the
condensation of bosons, which provides us a macroscopic counterpart
of relativistic particles described by the nonlinear Dirac equation
(NLDE). We first introduce the schemes and experiment progress on
creating synthetic gauge potentials in neutral atoms. Then we use
two specific systems to show that a liner and nonlinear Dirac
equation with tunable parameters through laser fields can be
simulated.

\subsection{Synthetic gauge fields for neutral atoms}

\noindent We first briefly review the generation of synthetic gauge
fields for neutral cold atoms loaded in OLs. In Section II.A.3, we
have mentioned that a U(1) Abelian gauge potential can be generated
by atom-laser interaction~[52] or temporal modulation of the lattice
potential~[53,54] for atoms trapped in a square OL. This external
gauge potential behaves as a vertical staggered or uniform (when
$\gamma$ is spatially independent) magnetic field acting on a
charged particle moving in a square potential, and thus contributes
a scalar Peierls phase factor to the atomic hopping process. A more
straightforward way to generate an Abelian gauge potential in OLs
can be achieved by using laser-assisted tunneling for atoms trapped
in state-dependent lattice~[71,73], where atoms also acquire a phase
when tunneling between the two neighbor sublattices. The phase
factor can extend to a $2\times 2$ matrix, corresponding to a square
lattice subjected to a U(2) non-Abelian gauge potential~[94]. The
basic idea to create a U(N) non-Abelian gauge potential in an OL is
to additionally consider atomic species with $N$ internal
quasi-degenerate sub-levels in both two sublattices~[72], and thus
instead of a simple phase, the laser-assisted tunneling would
induced a rotation in the internal space, described by a $N\times N$
matrix {\sl {\bf F}} with non-commuting component $F_x,F_y,F_z$.

An alternative method to create synthetic gauge field is rotating
the neutral atomic gas trapped in harmonic potentials or OLs~[106].
This scheme is limited to rotationally symmetric configurations, and
furthermore, can not extend to non-Abelian case without additional
laser fields~[107]. Other schemes that can be used to create a gauge
field include using light beams with orbital angular
momentum~[93-95] or spatially shift laser beams~[96-98] to interact
with atomic gases. Two reviews on the synthetic gauge potentials for
neutral atoms are given in Refs.~[99] and [108].

The relativistic Dirac Hamiltonian always includes a spin-orbit
coupling term (as shown in Eq. (5)), which can be induced from a
U(2) non-Abelian gauge potential. To achieve this goal without OLs,
we can consider the scheme based on the adiabatic motion of
multiple-level atoms in laser fields that generate (near-)degenerate
dark states~[100]. This idea goes back to the work by Wilczek and
Zee~[101], who have shown that non-Abelian gauge fields can arise in
the adiabatic evolution of quantum systems with multiple degenerate
eigenstates. We consider the adiabatic motion of atoms with $N+1$
internal levels in stationary laser fields, as shown in Fig. 9(a).
For fixed position $\mathbf{r}$ the internal Hamiltonian including
the laser-atom interaction can be diagonalized to yield a set of
$N+1$ dressed states $\left| \chi_n(\mathbf{r}) \right\rangle $ with
eigenvalues $\varepsilon _{n}(\mathbf{r})$, where $n=1,2,\ldots
,N+1$. The full quantum state of the atom describing both internal
and motional degrees of freedom can then be expanded in terms of the
dressed states according to $|\Phi\rangle =\sum_{n=1}^{N+1}\Psi
_{n}(\mathbf{r}) \left| \chi_n(\mathbf{r})\right\rangle$, where the
wavefunctions $\Psi=(\Psi_1,\Psi_2,\dots,\Psi_{N+1})^\top$ obeys the
Schr\"{o}dinger equation

\vspace{0.25cm} $\displaystyle i\hbar \frac{\partial}{\partial
t}\Psi =\left[\frac{1}{2m} (\mathbf{P} - \mathbf{A})^2 + V\right]
\Psi$ \hfill (28)

\vspace{0.25cm} \noindent with ${V}$ being an external trapping
potential. Here the vector potential $\mathbf{A}$ and scalar
potential $V$ are $(N+1)\times (N+1)$ matrices come from the spatial
dependence of the atomic dressed states with elements

\vspace{0.25cm} $\displaystyle \mathbf{A}_{n,m} = i\hbar \langle
\chi_n(\mathbf{r})| \nabla\chi_m(\mathbf{r}) \rangle$ ,\hfill (29a)

\vspace{0.15cm} $\displaystyle V_{n,m} = \varepsilon
_{n}(\mathbf{r})\, \delta_{n,m} +\langle
\chi_n(\mathbf{r})|V(\mathbf{r})|\chi_m(\mathbf{r})\rangle$.\hfill
(29b)

\vspace{0.25cm} \noindent If the off-diagonal elements of the
matrices $\mathbf{A}$ and $V$ are much smaller than the difference
of the dressed atomic energies, we can safely neglect the
off-diagonal contributions. In this regime, atoms in any one of the
dressed states adiabatically evolve according to a separate
Hamiltonian with a U(1) Abelian gauge potential, like a charged
particle moving in an electromagnetic field.

The key point to generate non-Abelian gauge potentials is that some
atomic dressed states, say the first $q$ ones, form a degenerate (or
nearly degenerate) manifold, leading to the failure of adiabatic
approximation since some off-diagonal coupling can no longer be
ignored. Suppose that these levels are well separated from the
remaining ones, the full Hamiltonian can be projected to this
subspace for the reduced column vector
$\tilde\Psi=\left(\Psi_1,\dots,\Psi_q\right)^\top$, which obeys the
Schr\"{o}dinger equation $i\hbar\frac{\partial}{\partial
t}\tilde\Psi={\tilde H}\tilde\Psi$ with the Hamiltonian

\vspace{0.25cm} $\displaystyle {\tilde H}= \frac{1}{2m} (\mathbf{P}
-\mathbf{A})^2 + V +\Phi  $,\hfill (30)

\vspace{0.25cm} \noindent where $\mathbf{A}$ and $V$ being the
truncated $q\times q$ matrices with elements defined in Eq. (29),
and a scalar potential $\Phi$ arises from the projection, with
elements given by

\vspace{0.25cm} $\displaystyle \Phi _{n,m}
=\frac{1}{2m}\sum_{l=q+1}^{N}\mathbf{A}_{n,l}\cdot
\mathbf{A}_{l,m}$. \hfill (31)


\vspace{0.25cm} \noindent  Here the effective U(q) non-Abelian gauge
potential $\bf{A}$ also called the Mead-Berry connection is related
to an effective magnetic field (or curvature) $\bf{B}$:


\vspace{0.25cm} $\displaystyle {\bf B}=\nabla\times {\bf
A}+\frac{1}{i\hbar}{\bf A} \times {\bf A} .$\hfill (32)

\vspace{0.25cm} \noindent Because the vector components of $\bf{A}$
do not necessarily commute, the term $\bf{A} \times \bf{A} $ does
not vanish in general, even if the vector potential is uniform in
space. In fact, this term reflects the non-Abelian character of the
gauge potentials and appears only if $q\geq 2$. When $q=2$
corresponding to a tripod scheme discussed in the following section,
under certain conditions the non-Abelian gauge structure is
equivalent to a spin-orbit interaction.

\vspace{3mm}
\includegraphics[width=7cm,height=7cm]{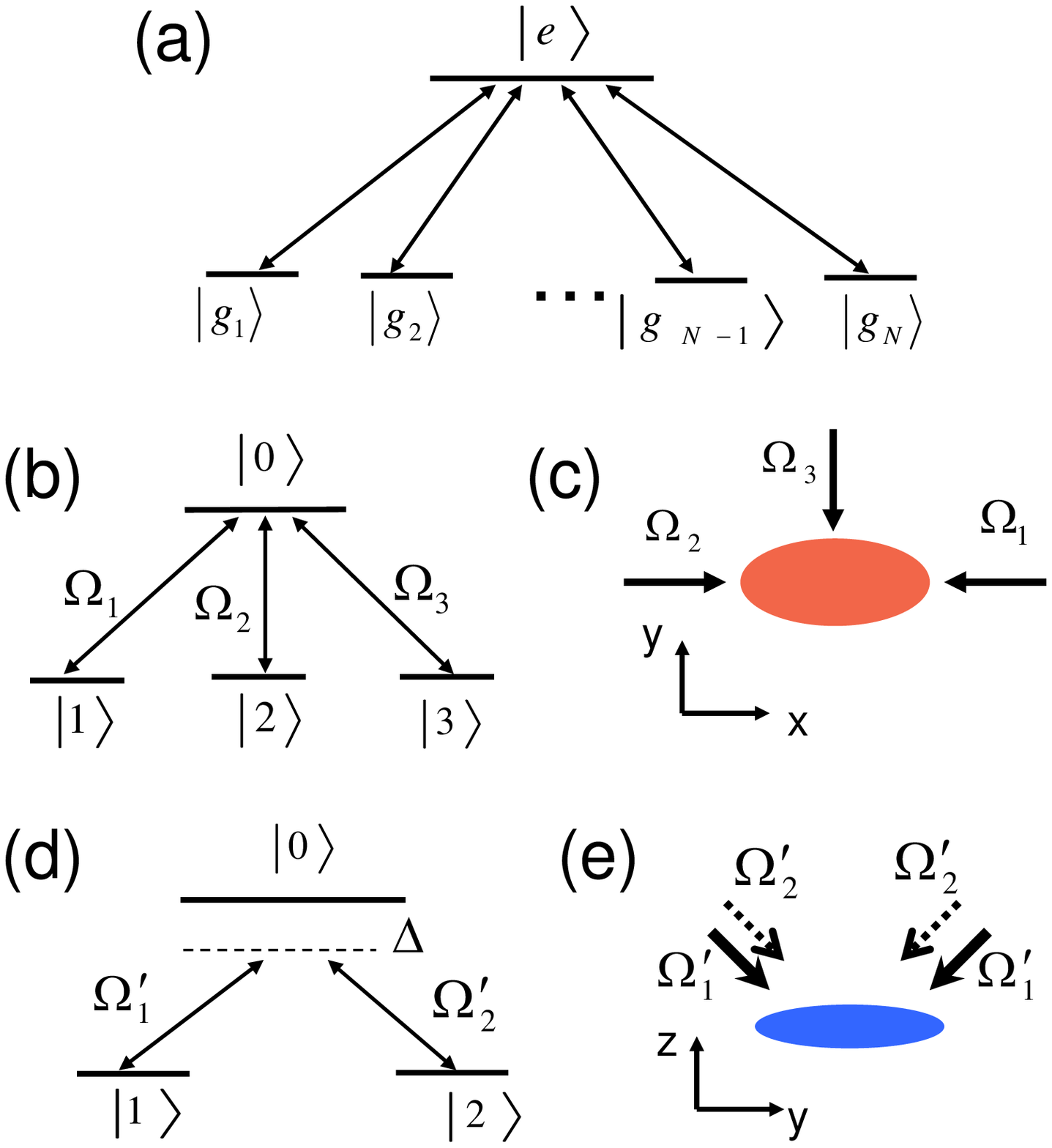} \vspace{1mm}
{\baselineskip
10.5pt\renewcommand{\baselinestretch}{1.05}\footnotesize

\noindent{\bf Fig.~9}\quad Schematic of atom-laser interaction. (a)
Multi-pod configuration. An atomic state $|e\rangle$ is coupled to
$N$ different atomic states $|g_j\rangle$ ($j=1,...,N$) by $N$
resonant laser fields. (b),(c) Atoms with tripod-level configuration
($N=3$) interacting with three laser beams characterized by the Rabi
frequencies $\Omega_1$, $\Omega_2$ and $\Omega_3$. (d),(e) Atoms
with $\Lambda$-level configuration interacting with laser beams
characterized by the Rabi frequencies $\Omega'_1$, $\Omega'_2$ and a
large detuning $\Delta$.

}\vspace{2mm}

The above single particle physics of generating gauge fields is
equivalent for bosonic and fermionic atoms. For a many-body system
of bosons, such as a BEC, the U(1) Abelian gauge potential acts like
a magnetic field and leads to a vortex lattice in the BEC~[98,109].
The NIST group has recently realized the Abelian gauge potential in
a BEC of $^{87}Rb$ atoms~[110,111]. In their experiments, they used
a gradient of the detuning of the laser frequency for the Raman
transitions between different atomic ground sub-levels to generate
an artificial gauge potential, somewhat different from the scheme we
discussed above. Besides, the NIST group has simulate an electric
force acting on neutral atoms through the time dependence of an
effective vector potential~[112]. A many-body system of
pseudo-spin-1/2 bosons within the tripod scheme was first considered
in Ref.~[113], and the degenerate condensation ground state of the
system was found in the presence of weak interactions. Furthermore,
in Refs.~[114,115], a spin-orbit coupled spinor BEC was investigated
and found to have two different phases in the ground state depending
on the interatomic interactions, a single plane wave phase with a
single dressed state and a standing wave with a nontrivial spin
stripe. Most excitingly, the NIST group in a very recent experiment
has reported the first realization of an effective U(2) non-Abelian
gauge potential in neutral atoms -- a spin-orbit coupled BEC~[116].

\subsection{Tunable Dirac-type equation for cold atoms with tripod- and $\Lambda$- level structure}

\noindent  Specializing the above scheme of generating U(N)
non-Ablelian gauge potential in neutral atoms, we show in this
section two different systems, where a synthetic U(2) non-Abelian
gauge field appears, can use to simulate the relativistic Dirac
equation.

{\sl Tripod-level configuration.---} Let us first consider the
adiabatic motion of atoms in $x$-$y$ plane with each having a
tripod-level structure in a laser field as shown in Fig. 9(b) and
(c)~[100,102]. The atoms in three lower levels $|1\rangle$,
$|2\rangle$ and $|3\rangle$ are coupled with an excited level
$|0\rangle$ through three laser beams characterized by the Rabi
frequencies $\Omega_1=\Omega\sin\theta\mathrm{e}^{-i\kappa
x}/\sqrt{2}$, $\Omega_2=\Omega\sin\theta\mathrm{e}^{i\kappa
x}/\sqrt{2}$ and $\Omega_3=\Omega\cos\theta\mathrm{e}^{-i\kappa y}$,
respectively, where
$\Omega=\sqrt{|\Omega_1|^2+|\Omega_2|^2+|\Omega_3|^2}$ is the total
Rabi frequency, and the mixing angle  $\theta$ defines the relative
intensity. The Hamiltonian of a single atom reads

\vspace{0.25cm} $\displaystyle
\hat{H}=\frac{\mathbf{P}^2}{2m}+V(\mathbf{r})+H_{int}$, \hfill (33)

\vspace{0.25cm} \noindent where $V=\sum_{j=1}^3 V_j(
\mathbf{r})|j\rangle\langle j|$ is the external trapping potential
and the laser-atom interaction Hamiltonian $H_{int}$ in the
interaction representation is

\vspace{0.25cm} $\displaystyle
H_{int}=-\hbar\Bigl(\Omega_1|0\rangle\langle1|+\Omega_2|0\rangle\langle2|
+\Omega_3|0\rangle\langle3|\Bigr)+\mathrm{H.c.}$ \hfill (34)

\vspace{0.25cm} \noindent Diagonalizing the Hamiltonian $H_{int}$
yields two degenerate dark states of zero energy

\vspace{0.25cm} \noindent$\displaystyle |D_1\rangle  =
\frac{1}{\sqrt{2}}e^{-i\kappa y}\left(e^{i\kappa
x}|1\rangle-e^{-i\kappa x}|2\rangle\right)$,\hfill (35a)

\vspace{0.15cm} \noindent$\displaystyle |D_2\rangle  =
\frac{1}{\sqrt{2}}e^{-i\kappa y}\cos\theta\left(e^{i\kappa
x}|1\rangle+e^{-i\kappa
x}|2\rangle\right)-\sin\theta|3\rangle$,\hfill (35b)

\vspace{0.25cm} \noindent as well as two bright state separated from
the dark states by the energies $\pm \hbar\Omega$. If $\Omega$ is
sufficiently large compared to the two-photon detuning due to the
laser mismatch and/or Doppler shift, then one can safely study only
the internal state of an atom adiabatically evolves within the dark
state manifold. In such a resonant coupling configuration, there is
atomic dissipation due to the spontaneous emission of the excite
state, thus the dark states acquire a finite lifetime. However, it
can be up to a few seconds for cold atoms with typical velocity
being of the order of a centimeter per second. The lifetime of the
dark states can be further increased by using off-resonant coupling
configuration as described below.

In the dark state manifold, the atomic state vector $|\Phi\rangle$
can be expanded in terms of the dark states according to
$|\Phi\rangle=\sum_{j=1}^2\Psi_j(\mathbf{r})|D_j(\mathbf{r})\rangle$,
where $\Psi_j(\mathbf{r})$ is the wave-function for the center-of-
mass motion of the atom in the $j$-th dark state. Thus the dynamics
of the atom is described by a two-component wavefunction
$\tilde\Psi=(\Psi_1,\Psi_2)^{T}$, which obeys the Schr\"odinger
equation (30) with $q=2$. The non-Abelian gauge potential
$\mathbf{A}$ in the present configuration of the light field is of
U(2) non-Abelian one. Substituting Eq. (35) into Eq. (29) and (31),
one can work out the expression of potentials $\mathbf{A}$, $\Phi$
and $V$:

\vspace{0.25cm} $\displaystyle \mathbf{A}  = \hbar\kappa\left(
\begin{array}{cc}
\mathbf{e}_y & -\mathbf{e}_x\cos\theta\\ -\mathbf{e}_x\cos\theta &
\mathbf{e}_y\cos^2\theta
\end{array}\right), \hfill (36a)
$

\vspace{0.15cm} $\displaystyle \Phi  = \left(
\begin{array}{cc}
\hbar^2\kappa^2\sin^2\theta/2m & 0\\ 0 &
\hbar^2\kappa^2\sin^2(2\theta)/8m
\end{array}\right)$, \hfill (36b)

\vspace{0.15cm} $\displaystyle V  = \left(
\begin{array}{cc}
V_1 & 0\\ 0 & V_1\cos^2\theta+V_3\sin^2\theta\end{array}\right)$,
\hfill (36c)

\vspace{0.25cm} \noindent where the trapping potential is assumed to
be the same for the first two atomic states, $V_1=V_2$. Choose
$V_1-V_3=\hbar^2\kappa^2\sin^2\theta/2m$, and thus the overall
trapping potential simplifies to $V+\Phi=V_1\mathrm{I}$ with
$\mathrm{I}$ being the unit matrix. Furthermore, let the mixing
angle $\theta=\theta_0$ to be such that
$\sin^2\theta_0=2\cos\theta_0$, giving $\cos\theta_0=\sqrt{2}-1$.
Thus the vector potential $\mathbf{A}$ takes a symmetric form

\vspace{0.25cm} $\displaystyle
\mathbf{A}=\hbar\kappa^{\prime}(-\mathbf{e}_x\sigma_x+\mathbf{e}_y\sigma_z)
+\hbar\kappa_0\mathbf{e}_y\mathrm{I}$ \hfill (37)

\vspace{0.25cm} \noindent in terms of the Pauli matrices $\sigma_x$
and $\sigma_z$, where $\kappa^{\prime}=\kappa\cos\theta_0$ and
$\kappa_0=\kappa(1-\cos\theta_0)$. Using a unitary transformation
$\tilde{H}'=U^{\dag}\tilde{H}U$ in Eq. (30) with
$U=\exp(-i\kappa_0y)\exp\left(-i\frac{\pi}{4}\sigma_x\right)$, one
obtains the Hamiltonian for the atomic motion

\vspace{0.25cm} $\displaystyle
\tilde{H}'=\frac{1}{2m}(\mathbf{P}+\hbar\kappa'\mathbf{\mathbf{
\sigma}_{\bot}})^2+V_1 $, \hfill (38)

\vspace{0.25cm} \noindent where
$\mathbf{\sigma}_{\bot}=\mathbf{e}_x\sigma_x+\mathbf{ e}_y\sigma_y$
is the spin $1/2$ operator in the $xy$ plane. In the limit of low
momenta, i.e., $|{\mathbf P}|\ll \hbar \kappa'$, one can safely
neglect the kinetic energy term in Eq. (38) and obtain an effective
Dirac Hamiltonian

\vspace{0.25cm} $\displaystyle H_{2D}=\hbar
c_0\mathbf{k}\cdot\mathbf{\sigma}_{\bot}+V_1+mc_0^2$, \hfill (39)

\vspace{0.25cm} \noindent where $c_0=\hbar\kappa'/m$ is the
effective light velocity for such quasi-relativistic particles. In
this case, the atoms in the system behave as 2D massless
relativistic particles.

If we focus on the case with an extremely anisotropic potential,
which is sufficiently strong along the transverse direction of the
$x$ axis, such that the original massive atoms are actually confined
in a one-dimensional guide along the $x$ axis with the vector
potential $A_x=-\hbar\kappa \cos\theta\sigma_x$. In this case, we
get a 1D Dirac equation described by~[103,104]

\vspace{0.25cm} $\displaystyle H_{1D} = c_{\star} \sigma_x p_x
+\gamma_z\sigma_z$, \hfill (40)

\vspace{0.25cm} \noindent up to an irrelevant constant, provided
that the wave vector of the atoms $ p_x /\hbar \ll \kappa
\cos\theta$. Comparing the original Dirac equation, here
$\gamma_z\equiv\hbar^2 \kappa^2\sin^4\theta/2m$ is the effective
rest energy and $c_{\star}=\hbar \kappa\cos\theta/m$ is the
effective light velocity, and thus the effective mass
$m_{\star}=\frac{m}{2}\tan^2\theta \sin^2\theta$. Furthermore, if we
take $V_1-V_3=\hbar^2\kappa^2\sin^2\theta/2m$ as discussed above, we
are able to obtain a massless 1D Dirac equation when $\gamma_z$ in
Eq. (40) vanishes. Note that the mass $m_{\star}$ of the simulated
Dirac particle is not the mass $m$ of the cold atom itself and it is
a remarkable feature that the mass term in the simulated Dirac-like
equation can be controlled by the laser beams, so one can simulate a
tunable Dirac equation with cold atoms.

{\sl $\Lambda$-level configuration.---} We now turn to consider the
adiabatic motion of atoms in $y$-$z$ plane with each having a
$\Lambda$-level structure interacting with laser beams as shown in
Fig. 9(d) and (e)~[105,117]. The ground states $|1\rangle$ and
$|2\rangle$ are coupled to the excited state $|3\rangle$ through
laser beams characterized respectively with the Rabi frequencies
$\Omega'_1=\Omega'\cos(\kappa_y y)e^{-i\kappa_z z}$ and
$\Omega'_2=\Omega'\sin(\kappa_y y)e^{i(\pi-\kappa_zz)}$, where
$\Omega'=\sqrt{|\Omega'_1|^2+|\Omega'_2|^2}$. Such Rabi frequencies
can be realized with two pairs of plane-wave lasers as shown in Fig.
9(e). For instance, $\Omega'_1$ can be realized with a pair of
lasers with $\frac{1}{2}\Omega' \exp[i(-\kappa_z z \pm \kappa_y y
)]$, while $\Omega'_2$ is achieved with $\frac{1}{2}\Omega'
\exp[i(\pm \pi/2-\kappa_z z \pm \kappa_y y)]$, where
$\kappa_y=\kappa\cos\varphi$ and $\kappa_z=\kappa\sin\varphi$ with
$\kappa$ being the wave number of the laser beams and $\varphi$
being the angle between the laser beams with the $y$ axis. Such
laser beams means that $\Omega'_1$ and $\Omega'_2$ are standing
waves in the $y$ direction but plane waves in the $z$ direction, and
both propagate opposite to the $z$ direction but the phase of
$\Omega'_1$ is $\pi$ ahead. The interaction Hamiltonian in Eq. (33)
in this configuration is given by

\vspace{0.25cm} $\displaystyle
H'_{int}=\hbar\Delta|3\rangle\langle3|-(\sum_{j=1}^2\hbar\Omega'_j|3\rangle\langle
j|+h.c.)$, \hfill (41)

\vspace{0.25cm} \noindent where $\Delta$ is the detuning.
Diagonalizing $H'_{int}$ yields a dark eigenstate
$|\chi^D\rangle=\sin(\kappa_y y)|1\rangle+\cos(\kappa_y y)|2\rangle$
and two bright eigenstates
$|\chi_1^B\rangle=(\sin\alpha)\cos(\kappa_y
y)|1\rangle-(\sin\alpha)\sin(\kappa_y y)|2\rangle-\cos\alpha
e^{-i\kappa_z z}|3\rangle$,
$|\chi_2^B\rangle=(\cos\alpha)\cos(\kappa_y y)|1\rangle-(\cos\alpha)
\sin(\kappa_y y)|2\rangle+(\sin\alpha) e^{-i\kappa_z z}|3\rangle$,
where $\alpha$ is determined by $\tan\alpha=
(\sqrt{\Delta^2+\Omega'^2}-\Delta)/\Omega'$. The corresponding
eigenvalues are $E^D=0$ and
$E_{1,2}^B=\hbar\left[\Delta\pm\sqrt{\Delta^2+4\Omega'^2}\right]/2$,
respectively. In the large detuning case $|\Delta| \gg \Omega'$, the
two lower eigenstates $|\chi^D\rangle$ and $|\chi^B_2\rangle$ span a
near-degenerate subspace, and have negligible contribution from the
state of $E^B_1$. Thus one can safely consider the adiabatic motion
of atoms in the near-degenerate subspace, leading to a U(2)
non-Abelian gauge potential with the vector and scalar potentials
obtained as $\mathbf{A}=-\frac{\hbar
\kappa_z\Omega'^2}{2\Delta^2}\sigma_z\vec{e}_z-\hbar
\kappa_y\sigma_y\vec{e}_y+\frac{\hbar
\kappa_z\Omega'^2}{2\Delta^2}\mathbf{I}\vec{e}_z$ and
$\Phi=\frac{\hbar^2\Omega'^2\kappa^2}{4m\Delta^2}[\cos(2\varphi)\sigma_z+\mathbf{I}]$
in this near-degenerate subspace with the basis $\{|\chi^D\rangle,
|\chi^B_2\rangle\}$. Under this condition, we obtain an effective
Hamiltonian

\vspace{0.25cm} $\displaystyle  \tilde{H}_k=
\frac{p_y^2+p_z^2}{2m}+v'_y\sigma_yp_y+v'_z\sigma_zp_z+\gamma'_z\sigma_z+V_T
$, \hfill (42)

\vspace{0.25cm} \noindent where the parameters $v'_y=\frac{\hbar
\kappa_y}{m}$, $v'_z=\frac{\hbar \kappa_z\Omega'^2}{2m\Delta^2}$,
and
$\gamma'_z=\frac{\hbar^2\Omega'^2}{4m\Delta^2}[\kappa_y^2-(1+\Omega'^2/\Delta^2)\kappa_z^2]+\frac{\hbar\Omega'^2}{2\Delta}+V_T$.
In the derivation, one has dropped an irrelevant constant and
assumed that the trapping potentials $V_{1,2,3}=V_T$ are
state-independent. The term related to $\sigma_zp_z$ in Eq. (42) can
be dropped approximately due to $v'_y\gg v'_z$. In fact, the
dynamics of atoms in the $z$ axis is still governed by the
Schr\"{o}dinger Hamiltonian since the kinetic energy term dominates,
and thus the atoms can be well confined by an optical waveguide
along $y$ axis~[103]. In the limit of low momenta, i.e., $p_y\ll
\hbar\kappa_y$, one can neglect the kinetic energy term. To this
step, one obtains the 1D effective Hamiltonian

\vspace{0.25cm} $\displaystyle H'_{1D}=
v'_y\sigma_yp_y+\gamma'_z\sigma_z+V_T $, \hfill (43)

\vspace{0.25cm} \noindent where $v'_y$ is the effective speed of
light in cold atoms and $\gamma'_z$ is the effective rest energy,
both of which are also controllable by the laser fields.

In the end of this section, we note that the relativistic
quasiparticles with total pseudospin $S=1$, which have been
described in Section II.A.2, can also be simulated with the tetrapod
($N=4$ in Fig. 9(a)) scheme~[118]. However, it looks like that the
present multi-pod coupling scheme is not feasible to simulate the
relativistic quasiparticles with higher (than $1$) or even arbitrary
pseudospin~[118]. To overcome this limitation, one may design other
atom-light interaction schemes which remain further work, or one can
adopt the proposal described in Ref.~[119], where the massless Dirac
fermions with arbitrary spin can be simulated by using
multicomponent ultracold atoms in a two-dimensional square OL.

\subsection{Simulation of nonlinear Dirac equation}

\noindent  In the previous section, we have shown that neutral cold
atoms in a light-induced non-Abelian gauge filed can be used to
simulate the 2D and 1D relativistic Dirac equation with tunable
parameters. Note that the above discussion is the single particle
physics and is suitable for both noninteracting bosons and fermions.
However, atomic interactions are natural and crucial in the atomic
assemble, such as BECs. Besides, the artificial non-Ableian gauge
potential has been realized in BECs~[116]. Thus one may ask whether
the relativistic quasiparticles can emergence in a many-body cold
atom system. In this section, we shall address that in the presence
of a laser-induced spin-orbit coupling an interacting ultracold BEC
may acquire a quasi-relativistic character described by a nonlinear
Dirac-like equation~[105,120].

In the tripod scheme~[120], one may choose another three laser beams
to get a more symmetrical gauge potential $\mathbf{A}=-\hbar\kappa
\sigma_y \vec{e}_x$. This may be achieved by employing three
co-propagating lasers along the z axis, with constant $|\Omega_3|$
and spatially dependent transversal profiles
$\Omega_1=|\Omega_3|\cos\phi(x,y)e^{i\kappa z}$,
$\Omega_2=|\Omega_3|\sin\phi(x,y)e^{i\kappa z}$ with a laser density
modulation along $x$ such that $\phi(\mathbf{r})=\sqrt{2}\kappa x$.
If the density modulation has however a polar symmetry on the $xy$
plane, i.e., $\phi(\mathbf{r})=\sqrt{2}\kappa \rho$ (with
$\rho^2=x^2+y^2$), then $\mathbf{A}=-\hbar\kappa \sigma_y
\vec{e}_\rho$, which is the 2D case. Finally, setting
state-dependent trapping potential
$V_1=V_2=\hbar\Delta_T-\hbar^2\kappa^2/2m$ with a detuning
$\Delta_T$, and $V_3=-V_1-2\hbar\Delta_T$, one obtains for both 1D
and 2D arrangements $\mathbf{\Phi}+V=\hbar\Delta_T\sigma_z$.

Interactions are characterized by $s$-wave scattering, and the
scattering lengths of different internal states are in principle
different, which may present collisionally induced spin rotations.
For simplicity of the discussion, we consider that the interaction
among the particles of different internal states are all identical,
with an effective $d$-dimensional interacting strength
$g=4\pi\hbar^2a_sN/(m\sqrt{2\pi}l_\bot^2)^{3-d}$, where $a_s$ is
scattering length, $N$ is the particle number, and $l_\bot$ is the
oscillator length associated to a harmonic vertical confinement. We
also assume that the interaction energy is much smaller than
$\hbar\Omega$, such that we remain in degenerate subspace.
Considering a wave-packet with $\langle p \rangle =\hbar k_0\ll
\hbar\kappa$ and momentum width $\delta p \ll 2\hbar\kappa$, we can
still safely neglect the $\mathbf{p}^2$ term. Within the
Gross-Pitaevskii formalism, the interacting bosons in the degenerate
subspace manifold are then effectively described by a NLDE as $
i\hbar\partial_t\Psi =H_{ND} \Psi$, where

\vspace{0.25cm} $\displaystyle
H_{ND}=\frac{\hbar\kappa}{m}\mathbf{p}\cdot{\mathbf
\sigma}+\hbar\Delta_T\sigma_z+g\Psi^\dag\cdot\Psi$. \hfill (44)

\vspace{0.25cm} \noindent Here the spin-orbit coupling term
$\mathbf{p}\cdot{\mathbf \sigma}=p_x\sigma_y$ for the 1D case, and
for the 2D case $\mathbf{p}\cdot{\mathbf \sigma}=p\sigma_y$ with
$p^2=p^2_x+p^2_y$.

For the $\Lambda$-level scheme~[105], the requirements of weak
atomic interactions and small momentum width for obtaining the NLDE
are similar with those in tripod scheme, i.e., the interaction
energy is much smaller than $\hbar\Omega'$ and $\delta
p_y\ll2\hbar\kappa_y$. In this configuration, the dynamics of a
spin-orbit coupled BEC are effectively described by the 1D NLDE with
the Hamiltonian

\vspace{0.25cm} $\displaystyle  H'_{ND}=-i\hbar
v'_y\sigma_y\partial_y+\gamma'_z\sigma_z+g\Psi^\dag\cdot\Psi+V_T$.
\hfill (45)

\vspace{0.25cm} \noindent Up to this, we have addressed that the
dynamics of a BEC with a light-induced spin-orbit coupling can
effectively described by a 1D or 2D NLDE under certain conditions.
Such quasi-relativistic BECs may become self-trapped and resemble
the so-called chiral confinement studied in the context of the
massive Thirring model~[120]. In Section IV.C, however, we will show
that macroscopic relativistic tunneling effect can be detected in
the $\Lambda$-scheme system.

Before closing this section, we note that the NLDE may also be
simulated with a BEC loading in the hexagonal optical lattice,
proposed in Refs.~[121-123]. It was suggested that the NLDE can be
obtained by cooling bosons to form a condensation in the lowest
Bloch band of a hexagonal optical lattice and then adiabatically
translate the BEC to the Dirac point at the band edge by
adiabatically tuning the relative phases between the laser
beams~[121,122]. In this system, the interplay between nonlinear
atomic interactions and Dirac dynamics may lead to rich localized
excitations~[122], including solitons, vortices, skyrmions, and
half-quantum vortices. However, it was recently argued that the
tight-binding model to describe such an interacting boson system
around the Dirac point is inadequate~[123]. To support this point,
it was showed that the atomic interaction of arbitrary small
strength can completely deform the topological structure of the
Bloch bands at the Dirac point~[123].

\section{Observation of some relativistic effects}

\noindent Some relativistic effects related to the Dirac equation,
have been predicted for a long time, the most well-known ones should
be the {\sl Zitterbewegung} (ZB) effect and the Klein tunneling
(KT). The concept of ZB was first introduced by
Schr\"{o}dinger~[39], who calculated the resulting time dependence
of the electron velocity and position in the framework of the Dirac
equation. He found that, in addition to classical motion, a free
relativistic electron exhibits a rapid periodic oscillations. The ZB
effect is due to an interference between positive and negative
energy state~[38], and will disappears with time if an electron is
represented by a wave packet since the interference has a finite
lifetime. Though Schr\"{o}dinger's idea stimulated numerous
theoretical investigation, direct experimental observation seems
impossible for free electrons since the predicted frequency
$\hbar\omega_Z\simeq2m_ec^2\simeq1$ MeV (here $m_e$ is the mass of
an electron) and the amplitude
$\lambda_c=h/m_ec\simeq3.86\times10^{-3}$ {\AA}.

Another exotic relativistic phenomena in the Dirac equation was
pointed out by Klein, who first used the Dirac equation to study an
electron scattering by a potential step and found that there exists
a nonzero transmission probability even though the potential hight
tends to infinity~[40], in contrast to the scattering of a
non-relativistic particle. This unique scattering process has
attracted lots of interest over the past eighty years but failed to
be directly tested by elementary particles due to the requirements
of currently unavailable electric field gradients, such as on the
order of $2m_ec^2/e\lambda_c\sim10^{18}$ V/m for free electrons.

In this section, we demonstrate that the famous ZB and KT can both
be observed in cold atom systems, where the dynamics of the
quasiparticles are described justly by the relativistic Dirac
equation. It is a remarkable feature that the effective mass and the
effective light speed are both controllable by the laser fields,
which makes the observation region can be achieved in experiments.
In principal, the OL systems discussed in Section II and the bulk
atom systems without OLs discussed in Section III, are both able to
exhibit the relativistic effects, we however here focus on the
latter for consistency but without loss of generality. We are also
interested in the tunneling issues of a spin-orbit coupled BEC
described by a NLDE, which may support the direct observation of
macroscopic KT under realistic conditions.

\subsection{{\sl Zitterbewegung}}

\noindent The concept of ZB was first rooted in the Dirac equation,
however, its present is not unique to the relativistic electrons,
but rather than a generic feature of spinor systems with Dirac-type
dispersion relationship. In solid state systems, the ZB can be
viewed as a special kind of inter-band transitions with generation
of virtual electron-hole pairs~[124]. The energy gaps in these
systems are usually of the order of meV and the oscillation
frequency is typically a few THz. Particularly, semiconductor
quantum wires~[125], graphene~[126,127], superconductors~[128], and
photonic crystals~[129,130], trapped ions~[10,131] have been
proposed as candidate systems for the observation of ZB. Excitingly,
a proof-of-principle quantum simulation of the ZB has been
respectively performed in the recent experiments in a photonic
crystal~[131] and trapped ion system~[10]. Since a Dirac-like
Hamiltonian can be created by cold atoms coupling with laser beams,
atoms in tripod-scheme system undergo ZB as a result~[104,132]. It
is also possible to realize ZB with cold atoms in an Abelian vector
potential~[133], and even to control the amplitude and lifetime of
ZB via an additional mirror oscillation~[134]. The ZB can also be
simulated with cold atoms in a driven~[135] or titled OL~[136]. In
these cold atomic systems, typically, the frequency of ZB is on the
order of a few MHz or kHz and the amplitude is comparable to the
wavelength of the laser beams. Thus they are promising candidates
for observing ZB. Here we give a complementary example to show the
atomic ZB by using the $\Lambda$- scheme.

\vspace{4mm}
\includegraphics[width=7.5cm,height=4cm]{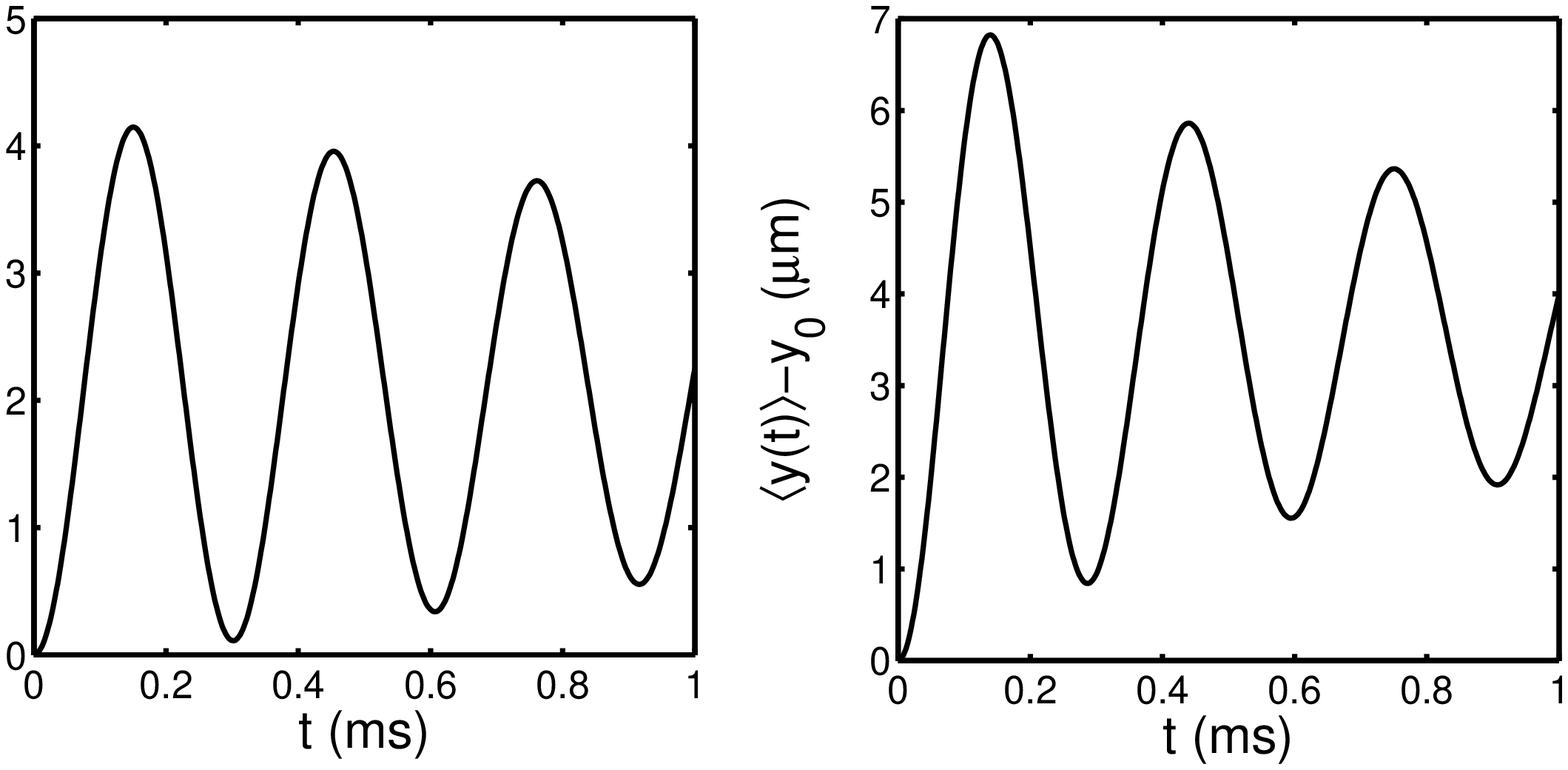} \vspace{1mm}
{\baselineskip
10.5pt\renewcommand{\baselinestretch}{1.05}\footnotesize

\noindent{\bf Fig.~10}\quad {\sl Zitterbewegung} for wave packets of
$^{7}$Li atomic gas with initial wave number $k_0 = 0$. (a)
$\kappa_y = 5.0\times10^6$ $m^{-1}$, (b) $\kappa_y = 1.0\times10^7$
$m^{-1}$. Other parameters are $m=1.16\times 10^{-26}$,
$\gamma'_z/\hbar = 10$ kHz and $\delta_l=10$ $\mu$m, respectively.

}\vspace{2mm}

In the $\Lambda$-scheme, the dynamics of atoms are effectively
described by the Dirac-like Hamiltonian (44). We consider the
Gaussian wave packet as the initial wave function

\vspace{0.25cm} $\displaystyle
\Psi(y,0)=\frac{1}{\sqrt{\delta_l\sqrt{\pi}}}e^{ik_0y}e^{-(y-y_0)^2/2\delta_l^2}\left(
                                                                        \begin{array}{c}
                                                                          c_1 \\
                                                                          c_2 \\
                                                                        \end{array}
                                                                      \right)
$ \hfill (46)

\vspace{0.25cm}\noindent in the spatial space, where $|c_i|^2$
($i=1,2$) normalized as $|c_1^2|+|c_2^2|=1$ determine the initial
population in states $|\chi_i\rangle$, $\delta_l$, $k_0$ and $y_0$
are initial wave-packet width, initial averages of wave number and
position, respectively. We suppose
$\delta_l=\sqrt{\hbar/m\omega_T}$, corresponding to the ground-state
width of the trapping potential $\frac{1}{2}m\omega^2_T y^2$ in
$y$-axis. In the momentum space, the initial wave packet is given by

\vspace{0.25cm} \noindent $\displaystyle
\Phi(k_y,0)=\frac{1}{\sqrt{2\pi}}\int\Psi(y,0)e^{-ik_yy}dy$

\vspace{0.15cm} $\displaystyle =\frac{1}{\sqrt{\delta_k\sqrt{\pi}}}
         \times e^{-i(k_y-k_0)y_0}e^{-(k_y-k_0)^2/2\delta_k^2}
         \left(
               \begin{array}{c}
               c_1 \\
               c_2 \\
               \end{array}
               \right)$, \hfill (47)

\vspace{0.25cm}\noindent  where $\delta_k=\delta_l^{-1}$ is the
momentum spread. When $t=0$, one turns off the trapping potential
$V_T$, and after $t$ time evolution governed by Dirac-type
Hamiltonian $H'_{1D}$ with $V_{T}=0$, the finial wave function is
written as

\vspace{0.25cm} $\displaystyle
\Psi(y,t)=\hat{\mathcal{T}}e^{-\frac{i}{\hbar}(v'_y\sigma_yp_y+\gamma'_z\sigma_z)t}\Psi(y,0)
$, \hfill (48)

\vspace{0.25cm}\noindent where $\hat{\mathcal{T}}$ denotes the time
ordering operator. It is straightforward to show its time evolution
$\Psi(y,t)=\frac{1}{\sqrt{2\pi}}\int\Phi(k_y,t)e^{ik_yy}dk_y$ with

\vspace{0.25cm} \noindent$\displaystyle
\begin{array}{llll}
\Phi(k_y,t)=\frac{1}{\sqrt{\delta_k\sqrt{\pi}}}
       e^{-i(k_y-k_0)y_0}e^{-(k_y-k_0)^2/2\delta_k^2}\\~~~~~
         \times \left(
               \begin{array}{rr}
               c_1\cos(\omega_kt)-\frac{v'_y k_y c_2+i\gamma'_z c_1/\hbar}{\omega_k}\sin(\omega_kt) \\
               c_2\cos(\omega_kt)+\frac{v'_y k_y c_1+i\gamma'_z c_2/\hbar}{\omega_k}\sin(\omega_kt) \\
               \end{array}
               \right)
\end{array}
$, \hfill (49)

\vspace{0.25cm}\noindent where
$\omega_k=\sqrt{(\gamma'_z/\hbar)^2+(v'_y k_y)^2}$ leads to
population transfer between two spin states, and thus $\Psi(y,t)$ is
sensitive to the initial spinor components. If we choose
$c_1=c_2=1/\sqrt{2}$, the wave packet of atomic gas undergoes
ZB~[104,130]. To investigate atomic ZB, we calculate the expectation
value of the center of mass, which is given by

\vspace{0.25cm} $\displaystyle \langle y(t)\rangle=i\int dk_y
\Phi^\dag(k_y,t)
\partial_{k_y} \Phi(k_y,t)$

\vspace{0.15cm} ~~~~~~$\displaystyle = y_0+\frac{v'_y
\gamma'_z}{\hbar\delta_k\sqrt{\pi}}\int dk_y \frac{\sin^2(\omega_k
t)}{\omega_k^2}e^{-(k_y-k_0)^2/\delta_k^2}$. \hfill (50)

\vspace{0.25cm}\noindent The integral term stands for ZB which shows
the oscillation of the center-of-mass motion. To have some ideas
about ZB, we numerically calculate Eq. (50) and two examples are
shown in Fig. 10. The reason for ZB is the separation of initial
wave packet by spin and then interfere between the coupling
components, which gives rise to oscillation of center-of-mass
motion. ZB oscillations for finite momentum spread damp out over
time, and its lifetime can be increases by reducing momentum spread
$\delta_k$. In this model, the amplitude of ZB is primarily
determined by the central momentum $k_0$ and the spin-orbit coupling
strength $v'_y$. The smaller $k_0$ and the larger $v'_y$, the larger
amplitude of ZB. The amplitude of ZB here can reach a few $\mu$m and
it is around $7$ $\mu$m when $k_0=0$, as shown in Fig. 10(b), which
is larger than that in early proposals on observations of ZB with
cold atoms~[104,132-134]. Meanwhile, the atomic ZB can persist for
several milliseconds with frequency being of the order of a few kHz.
ZB of atomic gas with such large amplitude and long lifetime is
detectable in current experiments.

\subsection{Klein tunneling}

\noindent The KT is originally referred to the step potential,
however, it has been extended to other kinds of potential barriers,
leading to a general description that relativistic particles can
penetrate through high and wide potential barriers without
exponential damping expected in non-relativistic tunneling
processes. Though this interesting relativistic scattering process
can not be directly tested by elementary particles, some
quasiparticles in certain systems may be described by effective
relativistic wave equation and thus provide a platform to simulate
the KT. Those including electrons in graphene~[137], electromagnetic
waves in honeycomb photonic crystals~[138], cold atoms in a
driven~[135] or bichromatic~[139,140] optical lattice, confined
stationary light~[141], and trap ions~[131], have been proposed to
observe such relativistic tunneling. Interestingly, in two very
recent experiments, a proof-of-principle simulation of KT with
trapped ions~[11] and with ultracold atoms in a bichromatic optical
lattice~[140] has been performed, respectively. We in the following
review a proposal of directly observing the KT with cold atoms in a
non-Abelian gauge field generated by using the
$\Lambda$-scheme~[105], as discussed in Section III.B, where we have
obtained a Dirac-like equation (43) with tunable parameters.

\vspace{3mm}
\includegraphics[width=8cm,height=2.5cm]{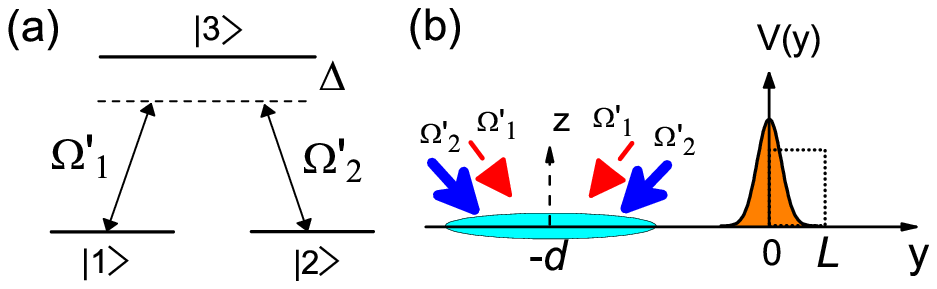} \vspace{1mm}
{\baselineskip
10.5pt\renewcommand{\baselinestretch}{1.05}\footnotesize

\noindent{\bf Fig.~11}\quad Schematic illustration of the system.
(a) Atom with $\Lambda$-level configuration interacting with laser
beams. (b) The configuration of the laser beams to realize a
Dirac-like equation and an effective Gaussian (square)-shape
potential induced by another laser beam. The atoms are confined in a
1D waveguide along $y$ axis and scattered by the potential.

}\vspace{2mm}

To have an intuitive physics picture, we first consider a single
atom with energy $E$ scattered by a square potential with the width
$L$ and potential hight $V_s$, as shown in Fig. 11(b). Such an
effective square potential can be experimentally formed by a laser
beam with flat-top profile~[142]. The transmission coefficient $T_D$
for the so-called KT regime is given by

\vspace{0.25cm} $\displaystyle T_D=\left[1+(\eta
-\eta^{-1})^2\sin^2(\alpha L)/4\right]^{-1}, $ \hfill (51)

\vspace{0.25cm}\noindent where
$\eta=\sqrt{\frac{(V_{\rm{s}}-E+\gamma'_z)(E+\gamma'_z)}{(E-V_{\rm{s}}+\gamma'_z)(\gamma'_z-E)}}$
and
$\alpha=\sqrt{\frac{(V_{\rm{s}}-E-\gamma'_z)}{(V_{\rm{s}}-E+\gamma'_z)}}/\hbar$.
Compared with the familiar property in the textbook of
non-relativistic quantum mechanics that the transmission coefficient
is a mono exponential decreasing as a function of the width $L$ or
the potential height $V_s$, a distinguished different feature within
this relativistic tunneling region is that the tunneling amplitude
can be an oscillation function of $V_s$ or $L$ even when the kinetic
energy of the incident particle is less than the height of the
square barrier potential. This relativistic effect can be attributed
to the fact that the incident particle in positive energy state can
propagate inside the barrier by occupying a negative energy state,
which is also a plane wave aligned in energy with that of the
particle continuum outside. Matching between positive and negative
energy sates across the barrier leads to the high-probability
tunneling. Furthermore, the resonant transmission occurs for perfect
matching where the potential width equates integral multiple of
half-wavelength of the negative energy state, corresponding to
$\alpha L=n\pi$ ($n=1,2,\cdots$) in the formula of $T_D$. We take
the atoms of $^{7}$Li with mass $m=1.16\times10^{-26}$ kg as an
example. If we choose the following typical experimental parameters:
the wave numbers $k_a=1.1\times10^6$ $\rm{m^{-1}}$, $\kappa_y=10^7$
$\rm{m^{-1}}$, $\kappa_z=0.8\times10^7$ $\rm{m^{-1}}$, the Rabi
frequency $\Omega=10^7$ Hz and the detuning $\Delta=10^9$ Hz, we can
find that the Klein regime corresponds to the required potential
height $V_s>0.162$ $\hbar\times$MHz, which can be easily achieved in
experiments. So we have demonstrated from a simple example that it
is feasible to observe the KT with cold atoms.

As for a practical experiment we are required to release two
conditions: the trajectory of a single atom is hard to detect, and
it is much easier in experiments to measure the density evolution of
an ensemble of noninteracting atoms. Compared with a perfect square
potential barriers achieved with a flat-top laser beams, a Gaussian
potential barrier $ V^{\rm{G}}_b(y)=V_{\rm{G}}e^{-y^2/\sigma^2}$,
where $V_{\rm{G}}$ is the height and $\sigma$ characterizes the
corresponding spatial variance, is much easier to be generated.
However, the conditions of resonant transmission varies with the
velocity and the width of the potential, and thus both the ensemble
of atoms and the Gaussian potential may smooth the oscillations in
the transmission coefficient.

So we now turn to address an ensemble of atoms that are scattered by
the Gaussian potential. We assume that the ensemble of atoms are
initially trapped in a harmonic trap which moves along the $y$ axis
with the wave number $k_a$. At the beginning $t=0$, the center of
the harmonic trap locates at $y=-d$, and the center of the Gaussian
potential barrier is at $y=0$, as shown in Fig. 11. In this case the
number density and the number wave number of the atomic ensemble are
characterized by Gaussian distribution. The trap is turned off at
$t=0$ and then we calculate the evolution of the density profile of
the atomic gas after a long enough time for scattering. We define
the average transmission coefficient for an ensemble of $N_a$
noninteracting atoms as

\vspace{0.25cm} $\displaystyle \langle
T\rangle=\frac{1}{N_a}\sum_{i=1}^{N_a}T(k_i),$ \hfill (52)

\vspace{0.25cm}\noindent where $T(k_i)$ denotes the transmission
coefficient of the atom $i$ scattered by the Gaussian potential, as
a function of random wave number $k_i$ described by the Gaussian
distribution $k_i \sim N(k_a,\sigma_k^2)$. The analytical expression
of $T(k_i)$ is absent, however, we here show an efficient method to
numerically solve it based on the transfer matrix methods. The
numerical procedures are outlined as follow. One first cuts the
Gaussian potential into spatially finite range $y\in[-y_c,y_c]$,
where the cutoff position $y_c$ should be chosen to guarantee that
the potential height outside the range is low enough to be
transparent for the atoms, i.e., $V_b^{\rm{G}}(y_c)\ll E, V_{\rm
G}$. Secondly, one equally divides this range into $n$ spindly
segments and each segment can be considered as a square potential if
$n$ is large enough. Thus the Gaussian potential can be
approximately viewed as a sequence of connective small square
potential barriers, and the transmission coefficient $T(k_i)\approx
1/|m_{11}|^2$, where $m_{11}$ is the first element of the whole
transfer matrix $M=M_nM_{n-1}\cdots M_j\cdots M_2M_1$. Here $M_j$
denotes the transfer matrix of the $j$-th square potential barrier,
whose explicit elements can be found in Ref.~[103]. Note that this
numerical calculation scheme recovers the non-relativistic
scattering governed by the Schr\"{o}dinger equation.

\vspace{3mm}
\includegraphics[width=8cm,height=4cm]{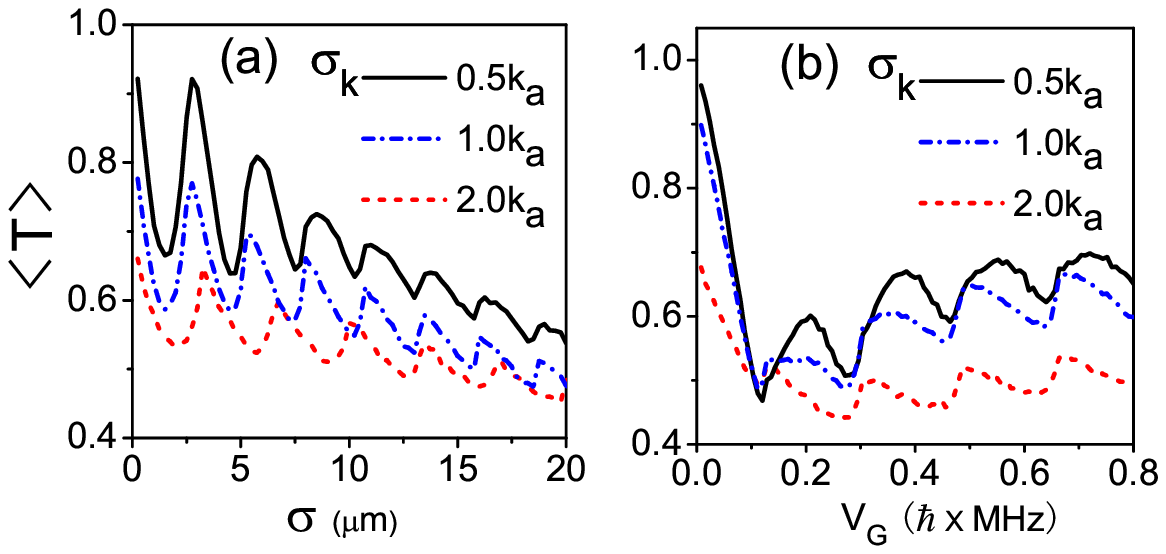} \vspace{1mm}
{\baselineskip
10.5pt\renewcommand{\baselinestretch}{1.05}\footnotesize

\noindent{\bf Fig.~12} \quad Klein tunneling of an ensemble of
$N_a=10^4$ noninteracting atoms. (a) $\langle T\rangle$ as a
function of potential width $\sigma$ with $\Omega_b^{\rm{G}}=0.8$
MHz. (b) $\langle T\rangle$ as a function of potential hight
$V_{\rm{G}}$ with $\sigma=10$ $\mu$m. The black solid line, blue
dashed line and red dotted line in both (a) and (b) correspond to
the cases of $\sigma_k=0.5,1.0,2.0 k_a$, respectively. Other
parameters are $k_a=1.1\times10^6$ $\rm{m^{-1}}$, $\kappa_y=10^7$
$\rm{m^{-1}}$, $\kappa_z=0.8\times10^7$ $\rm{m^{-1}}$, frequency
$\Omega=10^7$ Hz, and $\Delta=10^9$ Hz.

}\vspace{2mm}

Figure 12 shows the average transmission coefficient $\langle T
\rangle$ of an ensemble of $N_a=10^4$ noninteracting atoms as a
function of the width $\sigma$ and the height $V_G$ of the Gaussian
potential. Similar with the results for the single atom, the
tunneling oscillation still survives after the average of the random
distribution of the velocities for the Gaussian potential. The
amplitude of oscillation decreases with the increasing of the wave
number variance. Such KT features can be attributed to the fact that
the tunneling of an ensemble of noninteracting atoms is equivalent
to the superposition of every single-atom tunneling with different
oscillation period and amplitude, leading to smoothing the
oscillation property. The large velocity variance, the more obvious
smoothing effects. The oscillation period of single-atom tunneling
is insensitive to the potential hight in contract to the potential
width, and thus almost no decay of oscillation in Fig. 12(b). In
particular the interesting KT can be observed under realistic
conditions. For instance, the difference between the nearest peaks
of $\langle T \rangle $ in spatial dimension are in the region
$3.0-6.0\ \mu $m, and the frequency is about $0.2$ MHz. Both of them
in the oscillation are detectable within the current technology.

\subsection{Macroscopic Klein tunneling}

\noindent Although it is possible to observe the KT at a relatively
high temperature, the phenomena is clearer in low temperature. In
low temperature the bosonic atoms may form the BEC. Moreover, the
gauge field has recently been generated in a BEC by the NIST
group~[111,116], thus it deserves to study whether the KT is still
detectable with a BEC. Surprisingly we will illustrate below that
the KT of a spin-orbit coupled BEC may be observed very
clearly~[105].

The single-atom dispersion is characterized by two branches
$E_\pm(k_y)= \pm (\gamma_z'^2+\hbar^2v_y^2k_y^2)^{1/2}$, where the
lower (up) branch represents the negative (positive) energy state.
One can prepare an initial BEC with a designated mode $k_0$ at the
positive or negative energy branch. The two branches allow us to
investigate a more fruitful tunneling problem:  there are four
classes of the scattering which describe the wave function
$\Psi_\mu$ ($\mu (= \pm)$) scattered by the potential $\nu V_b{\rm
^{G}}$ with $\nu$ donating a barrier $(\nu=+)$ or a potential well
$(\nu=-)$, as shown in Fig. 13(a). We assume that the BEC is
initially trapped in a harmonic trap which moves along the $y$ axis,
thus we may choose the initial wave function of the BEC as

\vspace{0.25cm} $\displaystyle \Psi_\mu(y,0)=\frac{1}{\sqrt{
l_0\sqrt{\pi}}}e^{i\mu k_0y}e^{-(y+d)^2/2 l_0^2}\phi_\mu,$ \hfill
(53)

\vspace{0.25cm}\noindent where $l_0$ is the width, $k_0$ is the
central wave number of the wave-packet and the spinor $\phi_\mu$ are
defined as $ \phi_+ = ( i\cos\xi, -\sin\xi)^{T},$ $\phi_- =
(-i\sin\xi, \cos\xi )^{T}$ with $\xi=\frac{1}{2}\arctan(\hbar
v_yk_0/\gamma'_z)$ and $T$ as the transposition of matrix. Equation
(53) describes the Gaussian wave packet with the central velocity
$\hbar (\kappa_y+\mu k_0 )/m$ moving along $y$-axis. After the
evolution governed by Dirac-type (45) with time $t$ and replacing
$V_T$ by $\nu V^{\rm{G}}_b(y)$, the finial wave function becomes

\vspace{0.25cm} $\displaystyle \Psi_\mu(y,t)=\hat{\mathcal{T}}
\exp\left( -\frac{i}{\hbar}\int_0^t H'_{ND} dt\right)
\Psi_\mu(y,0).$ \hfill (54)

\vspace{0.25cm}\noindent We have numerically calculated
$\Psi_\mu(y,t)$ and as an example being shown in Fig. 13(b), the
stationary solution of the tunneling is always found within a few
milliseconds, which is a small time scale comparing with the
lifetime of BECs. After tunneling, the incident wave packet divides
into the left- and right-traveling wave packets and only the latter
one is on the transmission side of the barrier. Thus we can define
the transmission coefficient of the incident wave packet
$\Psi_\mu(y,0)$ scattering by a potential $\nu V_b{\rm ^{G}}$ as

\vspace{0.25cm} $\displaystyle  T_{\mu\nu} =
 \int_{\sigma}^\infty
\Psi_\mu^\dag(y,\tau)\Psi_\mu(y,\tau)dy,$ \hfill (55)

\vspace{0.25cm}\noindent where $\tau$ (being slightly larger than
$d/v_0$) represents a time that the reflected and transmitted wave
packets are sufficient away from the Gaussian potential. One can
directly measure the transmission coefficient in Eq. (55) since the
spatial density distribution $\rho_\mu
(y,\tau)=|\Psi_\mu(y,\tau)|^2$ can be detected using absorption
imaging~[143]. We first look into the tunneling phenomena for a
non-interacting BEC ($g=0$ in Eq. (45)) realized by Feshbach
resonance~[143], and then briefly discuss the effects of the
interaction between the atoms.

We plot the transmission coefficient  $T_{++}$ as a function of the
potential height $V_{\rm{G}}$ and width $\sigma$ in Fig. 14(a,b)
with the practical experimental parameters. It is interesting to
note that the transmission coefficient decreases exponentially to
zero with $V_{\rm{G}}$ when $V_{\rm{G}}<V_{\rm{G}}^K$, while if we
further increase the potential height to the Klein region
$V_{\rm{G}}>V_{\rm{G}}^K$, the transmission coefficient will
increase and then be an oscillating function in the Klein region
$V_{\rm{G}}>V_{\rm{G}}^K$, being similar to the results of Eq. (51)
for the which describes the transmission of the square potential
barrier. Here the critical value of the potential height may
approximately be estimated using the square potential barrier with
$V_{\rm{G}}^K= E(k_0)+\gamma_z \approx 0.09$ MHz. Besides, there are
two identities $T_{++}=T_{--}$ and $T_{-+}=T_{+-}$ since Eq. (45)
with $g=0$ is invariant under the charge conjugation~[144], the
former as an example is confirmed in Fig. 14(a). Although the
amplitude of tunneling oscillation is less than the unit as
comparing with the tunneling of single atom, the amplitude of
tunneling oscillation can be more than $0.5$ and meanwhile the
period can be a few micrometers as shown in Fig. 14(b), which is
experimentally detectable.

The most exotic feature induced by the relativistic effects is that,
the BEC with negative energy can almost completely transmit a wide
Gaussian potential barrier, as shown in Fig. 14(c). This phenomena
can be attributed to the fact that the incident BEC in a negative
energy  level can be considered as a macroscopic 'anti-BEC'.
Alternatively, it can be understood that such scattering feature is
actually equivalent to that of a BEC of positive energy scattered by
a Gaussian potential well because of $T_{-+}=T_{+-}$. We also
calculate the transmission coefficient for the central mode of the
wave packet, as shown in the insert of Fig. 14(c), which further
confirms that a wide enough Gaussian potential well is transparent.
The reason lies in the fact that, in contract to square potential
wells, the Gaussian potential wells are smooth (without any energy
jump) in the whole space, and may support adiabatic motions of wave
packets in the large width limitation.

\vspace{3mm}
\includegraphics[width=7cm,height=3cm]{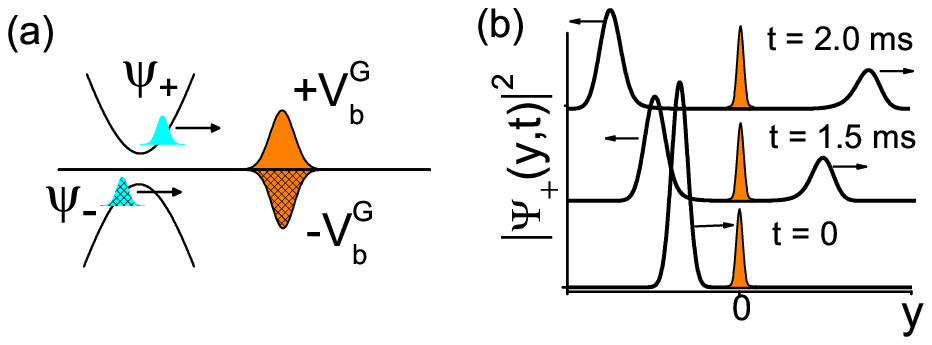} \vspace{1mm}
{\baselineskip
10.5pt\renewcommand{\baselinestretch}{1.05}\footnotesize

\noindent{\bf Fig.~13} \quad Schematic representation of the
scattering of a BEC. (a) A schematic diagram shows four kinds of
scattering events. (b) Normalized density distribution in a
scattering process at time $t=0, 1.5$ and $2.0$ ms. The peaks at
$y=0$ are the Gaussian barriers.

}\vspace{2mm}

The exotic tunneling property exhibited in Fig. 14(c) is an
intrinsic relativistic and macroscopic quantum phenomenon that can
not be explained with an incoherent ensemble average of a large
number of atoms. To clarify this point we also calculate numerically
the average transmission coefficient for an ensemble of $10^4$
noninteracting atoms as $\langle T\rangle$. Here we choose the wave
number distribution to be the same Gaussian distribution as that in
Eq. (53), i.e., $k_i \sim N(k_0,\sigma_k^2)$ with the variance
$\sigma_k=1/l_0$. The average coefficient $\langle T \rangle$ is
shown in the inset of Fig. 14(c), which is almost the same as that
of a single atom since $\sigma_k$ is small. The differences between
$\langle T \rangle $ and $T_{-+}$ in Fig. 14(c) clearly demonstrate
that the tunneling of BEC is not equivalent to an ensemble average
of the individual atoms with the same distribution of wave number.

Before ending this section, we make two additional comments. (i) The
quadratic term of the momentum in Eq. (42) has been neglected in the
derivation of the Dirac Eq. (45). To judge the feasibility of this
approximation, the transmission coefficients $T_{++}$ with or
without the quadratic term are compared in Fig. 14(b). It is shown
that the quadratic term leads to merely a slight left-shift of the
tunneling peaks. This  phenomenon can be interpreted by the fact
that  the wavelength of the 'anti-particle' consisting of the BEC
inside the barrier decreases slightly in the presence of the
additional low kinetic energy. This result verifies that the
approximation which leads to the effective Dirac equation is well
satisfied. (ii) We have also calculated the transmission coefficient
for the weak atomic interaction in BECs in Fig. 14(b), which shows
that the effect of the weak interaction  is little and smooths
merely the tunneling oscillation slightly. Therefore the exotic
tunneling phenomena addressed here survive in the case of weak
interaction between atoms.

\vspace{3mm}
\includegraphics[width=8cm,height=6cm]{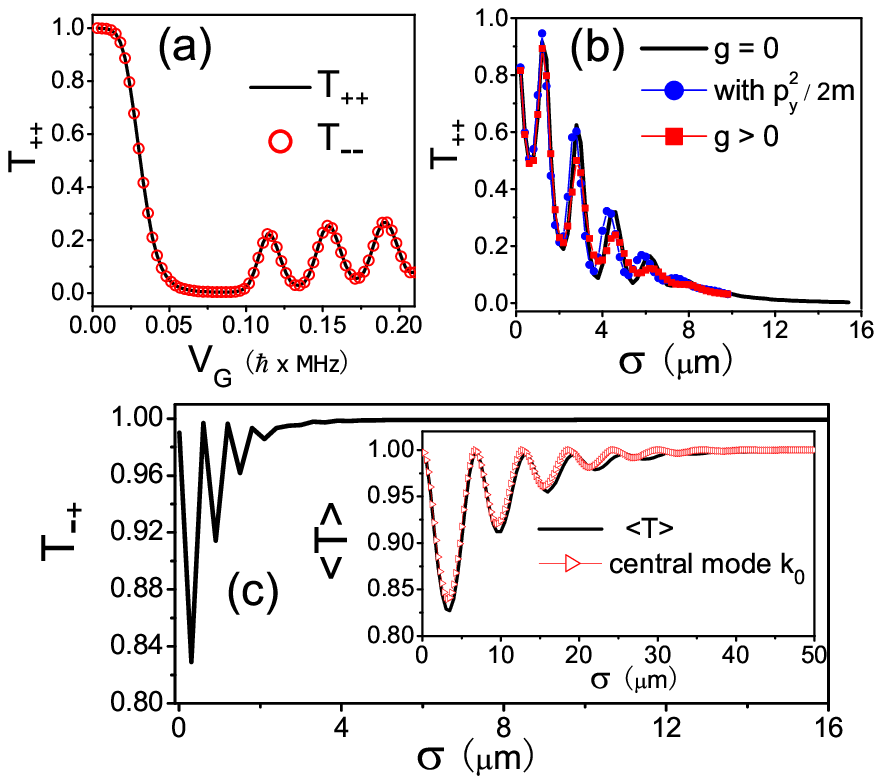} \vspace{1mm}
{\baselineskip
10.5pt\renewcommand{\baselinestretch}{1.05}\footnotesize

\noindent{\bf Fig.~14} \quad Klein tunneling of BECs. (a)  $T_{++}$
as a function of the potential height $V_G$ with $\sigma=5$ $\mu$m.
The relation $T_{++}=T_{--}$ is confirmed. (b) $T_{++}$ as a
function of potential width $\sigma$ with $V_{\rm{G}}=0.2$ MHz.
 The tunnelings of a BEC with the classic kinetic energy
term and the weak atomic interaction ($N = 2 \times 10^4$, $l_\perp
= 1.4$ $\mu$m and $a_s = 5a_0$ with $a_0$ being the Bohr radius) are
also depicted. (c) $T_{-+}$ as a function of potential width
$\sigma$ with $V_{\rm{G}}=0.2$ MHz. In the insert, the transmission
coefficient of a plane wave at {bf the} central mode $k_0$ is shown
as the line with labels of triangle, while the average transmission
coefficient $\langle T\rangle$ of $10^4$ atoms is shown as the solid
line. The other parameters: $l_0=10$ $\mu$m, $k_0=5.5\times10^5$
$\rm{m^{-1}}$, $\gamma_z/\hbar = 30$ kHz, and $d=4(l_0+\sigma)$.

}\vspace{2mm}

\section{Conclusions and perspectives}

\noindent In summary, we have reviewed recent progress on quantum
simulation of the relativistic Dirac equation by using ultracold
neutral atoms. As we have shown, ultracold atoms in a wide range of
systems, including the optical lattices with designated structures
and different light-induced gauge fields, can behave as the
relativistic particles under certain conditions. Comparing with
other relativistic systems such as the graphene, cold atom systems
offer rather more degrees of freedom to the relativistic
quasiparticles, which can exhibit from one-dimension to
three-dimension, and even from massive to massless. In other words,
one can simulate a highly tunable Dirac equation with cold atoms by
well dressing laser beams. Furthermore, the controllable atomic
interactions and disorders in these systems provide a unique
platform to investigate the relativistic dynamics and scattering
processes such as the celebrated {\sl Zitterbewegung} effect and
Klein tunneling as well as its extended macroscopic version.

Remarkably, some crucial experimental techniques for cold atoms used
to be Dirac equation simulators have been recently demonstrated. The
important progress includes addressing atomic gases in a hexagonal
optical lattice~[69], and generating an effective U(2) non-Abelian
gauge potential in a BEC~[116]. The latter technique is principally
available for fermionic atoms, which may provide an opportunity for
the realization of the long-sought Majorana fermions (see
Refs.~[145,146] and the referred works therein) in ultracold atomic
superfluid~[147-149]. The Majorana fermions described by the
Majorana equation are distinguished from the Dirac fermions by the
most exotic property that Majorana fermions are their own
antiparticles. The Majorana fermions with the feature of non-Abelian
statistics has raised significant interest as candidates for the
realization of topological quantum computation~[150]. To achieve
this goal, we will still face the arduous task of manipulating
Majorana fermions in cold atomic superfluid as that in other
proposed systems~[151]. Another important issue beyond the hunt for
Majonana fermions should be the quantum simulation of Majoranna
equation~[152], which is central to the recent work not only in
particle physics, supersymmetry and dark matter, but also some
exotic states of ordinary matter. If succeed, we are allowed to
explore some new exotic quantum relativistic processes that may go
beyond ordinary Dirac quantum mechanics.

In the present proposals of quantum simulation of the Dirac
equation, the atomic interactions are less considered due to the
complexity such as the nonlinear interacting for bosonic atoms.
Thus, it is of interest to ask whether and how the whole physics
picture of relativistic dynamics in the present of controllable
nonlinearity, that is an open question in the original Dirac theory,
can be learned by cold atom simulators~[105,120]. Besides, combining
with the controllable laser-formed disorder potential techniques, we
may also be able to investigate the Anderson localization of
relativistic particles~[103], especially the relativistic version of
competition between the interaction and disorder, which is still a
controversial problem even for the non-relativistic particles.
Furthermore, shall we be able to realize the quantum simulation of
interacting relativistic quantum field theories~[153] or find some
other exotic high-energy phenomena in these oppositely
lowest-temperature systems among the universe?

\vspace{1cm} {\bf Acknowledgements:} This work was supported by the
National Natural Science Foundation of China (Grant Nos. 10974059
and 11125417), the State Key Program for Basic Research of China
(Grant No. 2011CB922104), the GRF and CRF of the RGC of Hong Kong.


\begin{thebibliography}{99}

\bibitem{1.} R. Feynman, Int. J. Theor. Phys., 1982, 21: 467

\bibitem{2.} See the reviews in Insight: Quantum Coherence, Nature,
2008, 453: 1003-1049

\bibitem{3.} I. Buluta and F. Nori, Science, 2009, 326: 108


\bibitem{4.} M. Greiner, O. Mandel, T. Esslinger, T. W. H\"{a}nsch,
I. Bloch, Nature, 2002, 415: 39


\bibitem{5.} T. St\"{o}ferle, H. Moritz, C. Schori, M. K\"{o}hl, and
T. Esslinger, Phys. Rev. Lett., 2004, 92: 130403


\bibitem{6.} I. B. Spielman, W. D. Phillips, and J.V. Porto, Phys.
Rev. Lett., 2007, 98: 080404

\bibitem{7.} J. Simon, W. S. Bakr, R. Ma, M. E. Tai, P. M. Preiss,
and M. Greiner, Nature, 2011, 472: 307

\bibitem{8.}  A. Friedenauer, H. Schmitz, J. T. Glueckert, D.
Porras, and T. Schaetz, Nature Phys., 2008, 4: 757

\bibitem{9.} K. Kim, M. S. Chang, S. Korenblit, R. Islam, E. E.
Edwards, J. K. Freericks, G. D. Lin, L. M. Duan, and C. Monroe,
Nature, 2010, 465: 590

\bibitem{10.} R. Gerritsma, G. Kirchmair, F. Z\"{a}hringer, E.
Solano, R. Blatt, and C. F. Roos, Nature, 2010, 463: 68

\bibitem{11.} R. Gerritsma, R. Gerritsma, B. P. Lanyon, G.
Kirchmair, F. Z\"{a}hringer, C. Hempel, J. Casanova, J. J.
Garc\'{i}a-Ripoll, E. Solano, R. Blatt, and C. F. Roos, Phys. Rev.
Lett., 2011, 106: 060503

\bibitem{12.} J. T. Barreiro, M. M\"{u}ller, P. Schindler, D. Nigg,
T. Monz, M. Chwalla, M. Hennrich, C. F. Roos, P. Zoller, and R.
Blatt, Nature, 2011, 470: 486

\bibitem{13.} J. Du, N. Xu, X. Peng, P. Wang, S. Wu, and D. Lu,
Phys. Rev. Lett., 2010, 104: 030502

\bibitem{14.} M. H. Anderson, J. R. Ensher, M. R. Matthews, C. E.
Wieman, and E. A. Cornell, Science, 1995, 269: 198

\bibitem{15.} F. Dalfovo, S. Giorgini, L. P. Pitaevskii, and S.
Stringari, Rev. Mod. Phys., 1999, 71: 463

\bibitem{16.} C. A. Regal, M. Greiner, and D. S. Jin, Phys. Rev.
Lett., 2004, 92: 040403

\bibitem{17.} S. Giorgini, L. P. Pitaevskii, and S. Stringari, Rev.
Mod. Phys., 2008, 80: 1215

\bibitem{18.} J. Liu and B Liu, Front. Phys. China, 2010, 5: 123

\bibitem{19.} H. Jing, Y. Jiang, and Y. Deng, Front. Phys., 2011, 6:
15

\bibitem{20.} M. Lewenstein, A. Sanpera, V. Ahufinger, B. Damski, A.
Sende, U. Sen, Adv. Phys., 2007, 56: 243

\bibitem{21.} H. Zhai, Front. Phys. China, 2009, 4: 1

\bibitem{22.} A. Klein and D. Jaksch, Phys. Rev. A, 2006, 73: 053613

\bibitem{23.} M. A. Baranov, K. Osterloh, and M. Lewenstein, Phys.
Rev. Lett., 2005, 94: 070404

\bibitem{24.} A. S. S{\o}rensen, E. Demler, and M. D. Lukin, Phys.
Rev. Lett., 2005, 94: 086803

\bibitem{25.} G. B. Jo, Y. R. Lee, J. H. Choi, C. A. Christensen, T.
H. Kim, J. H. Thywissen,  D. E. Pritchard, and W. Ketterle, Science,
2009, 325: 1521

\bibitem{26.} J. Billy, V. Josse, Z. Zuo, A. Bernard, B. Hambrecht,
P. Lugan, D. Cl\'{e}ment, L. Sanchez-Palencia, P. Bouyer, and A.
Aspect, Nature, 2008, 453: 891

\bibitem{27.} G. Roati, C. D¡¯Errico, L. Fallani, M. Fattori, C.
Fort,
 M. Zaccanti, G. Modugno, M. Modugno, and M. Inguscio, Nature, 2008,
453: 895

\bibitem{28.} B. Deissler, M. Zaccanti, G. Roati, C. D'Errico, M.
Fattori, M. Modugno, G. Modugno, and M. Inguscio,  Nature Phys.,
2010, 6: 354

\bibitem{29.} L. Sanchez-Palencia and M Lewenstein, Nature Phys.,
2010, 6: 87

\bibitem{30.} L. J. Garay, J. R. Anglin, J. I. Cirac, and P. Zoller,
Phys. Rev. Lett., 2000, 85: 4643

\bibitem{31.} U. R. Fischer and R. Sch\"{u}tzhold, Phys. Rev. A,
2004, 70: 063615

\bibitem{32.} M. Snoek, M.Haque, S. Vandoren, and H. T. C. Stoof,
Phys. Rev. Lett., 2005, 95: 250401

\bibitem{33.} Y. Yu and K. Yang, Phys. Rev. Lett., 2008, 100: 090404

\bibitem{34.} Y. Yu and K. Yang, Phys. Rev. Lett., 2010, 105: 150605

\bibitem{35.} K. S. Novoselov, A. K. Geim, S. V. Morozov, D. Jiang,
Y. Zhang, S. V. Dubonos, I. V. Grigorieva, and A. A. Firsov,
Science, 2004, 306: 666

\bibitem{36.} K. S. Novoselov, A. K. Geim, S. V. Morozov, D. Jiang,
M. I. Katsnelson, I. V. Grigorieva, S. V. Dubonos, and A. A. Firsov,
Nature, 2005, 438: 197

\bibitem{37.} M. Z. Hasan, C. L. Kane, Rev. Mod. Phys., 2010, 82:
3045

\bibitem{38.} W. Greiner, Relativistic Quantum Mechanics, 3rd Ed.,
Berlin: Spinger-Verleg, 2003

\bibitem{39.} E. Schr\"{o}dinger, Sitzungsber. Preuss. Akad. Wiss.
Phys. Math. Kl., 1930, 24: 418

\bibitem{40.} O. Klein, Z. Phys., 1929, 50: 157

\bibitem{41.} S. L. Zhu, B. Wang, and L. M. Duan, Phys. Rev. Lett.,
2007, 98: 260402

\bibitem{42.} E. Zhao and A. Paramekanti, Phys. Rev. Lett., 2006,
97: 230404

\bibitem{43.} C. Wu and S. Das Sarma, Phys. Rev. B, 2008, 77: 235107

\bibitem{44.} B. Wunsch, F. Guinea, and F. Sols, New J. Phys., 2008,
10: 103027

\bibitem{45.} K. L. Lee, B. Gr\'{e}maud, R. Han, B. G. Englert, and
C. Miniatura, Phys. Rev. A, 2009, 80: 043411

\bibitem{46.} A. Dutta, R. R. P. Singh, and U. Divakaran, Europhys.
Lett., 2010, 89: 67001

\bibitem{47.} D. Poletti, C. Miniatura, and B. Gr\'{e}maud,
Europhys. Lett., 2011, 93: 37008

\bibitem{48.} D. Bercioux, D. F. Urban, H. Grabert, and W.
H\"{a}sler, Phys. Rev. A, 2009, 80: 063603

\bibitem{49.} D. Bercioux, N. Goldman, and D. F. Urban, Phys. Rev.
A, 2011, 83: 023609

\bibitem{50.} R. Shen, L. B. Shao, B. Wang, and D. Y. Xing, Phys.
Rev. B, 2010, 81: 041410(R)

\bibitem{51.} I. I. Satija, D. C. Dakin, J. Y. Vaishnav, and C. W.
Clark, Phys. Rev. A, 2008, 77: 043410

\bibitem{52.} J. M. Hou, W. X. Yang, and X. J. Liu, Phys. Rev. A,
2009, 79: 043621

\bibitem{53.} L. K. Lim, C. M. Smith, and A. Hemmerich, Phys. Rev.
Lett., 2008, 100: 130402

\bibitem{54.} L. K. Lim, A. Hemmerich, and C. M. Smith, Phys. Rev.
A, 2010, 81: 023404

\bibitem{55.} L. K. Lim, A. Lazarides, A. Hemmerich, and C. M.
Smith, Europhys. Lett., 2009, 88: 36001

\bibitem{56.} N. Goldman, A. Kubasiak, A. Bermudez, P. Gaspard, M.
Lewenstein, and M. A. Martin-Delgado, Phys. Rev. Lett., 2009, 103:
035301

\bibitem{57.} X. J. Liu, X. Liu, C. Wu, and J. Sinova, Phys. Rev. A,
2010, 81: 033622

\bibitem{58.} M. P. Kennett, N. Komeilizadeh, K. Kaveh, and P. M.
Smith, Phys. Rev. A, 2011, 83: 053636

\bibitem{59.} M. Yang and S. L. Zhu, Phys. Rev. A, 2010, 82: 064102

\bibitem{60.} A. Bermudez, L. Mazza, M. Rizzi, N. Goldman, M.
Lewenstein, and M. A. Martin-Delgado, Phys. Rev. Lett., 2010, 105:
190404

\bibitem{61.} L. Lepori, G. Mussardo, and A. Trombettoni, Europhys.
Lett., 2010, 92: 50003

\bibitem{62.} L. M. Duan, E. Demler, and M. D. Lukin, Phys. Rev.
Lett., 2003, 91: 090402

\bibitem{63.} Y. Zhang, Y. Tan, H. L. Stormer, and P. Kim, Nature,
2005, 438: 201

\bibitem{64.} G. W. Semenoff, Phys. Rev. Lett., 1984, 53: 2449

\bibitem{65.} X. G. Wen, Quantum Field Theory of Many-Body Systems,
Oxford: Oxford University, 2004.

\bibitem{66.} A. Bermudez, N. Goldman, A. Kubasiak, M. Lewenstein,
and M. A. Martin-Delgado, New J. Phys., 2010, 12: 033041

\bibitem{67.} J. K. Block and N. Nygaard, Phys. Rev. A, 2010, 81:
053421

\bibitem{68.} C. Wu, D. Bergman, L. Balents, and S. Das Sarma, Phys.
Rev. Lett., 2007, 99: 070401

\bibitem{69.} P. Soltan-Panahi, J. Struck, P. Hauke, A. Bick, W.
Plenkers, G. Meineke, C. Becker, P. Windpassinger, M. Lewenstein and
K. Sengstock ,  Nature Phys., 2011, 7: 434

\bibitem{70.} V. Apaja, M. Hyrk\"{a}s, and M. Manninen,  Phys. Rev.
A, 2010, 82: 041402(R)

\bibitem{71.} D. Jaksch and P. Zoller, New J. Phys., 2003, 5: 56

\bibitem{72.} K. Osterloh, M. Baig, L. Santos, P. Zoller, and M.
Lewenstein, Phys. Rev. Lett., 2005, 95: 010403

\bibitem{73.} F. Gerbier and J. Dalibard, New J. Phys., 2010, 12:
033007

\bibitem{74.} N. Goldman, A. Kubasiak, P. Gaspard, and M.
Lewenstein, Phys. Rev. A, 2009, 79: 023624

\bibitem{75.} L. H. Karsten and J. Smith, Nucl. Phys. B, 1981, 183:
103

\bibitem{76.} H. B. Nielsen and M. Ninomiya, Nucl. Phys. B, 1981,
185: 20

\bibitem{77.} K. Wilson, New Phenomena in Subnuclear Physics.,
edited Plenum: New York, 1977

\bibitem{78.} X. L. Qi, R. Li, J. Zang, S. C. Zhang, Science, 2009,
323: 1184

\bibitem{79.} M. W. Zwierlein, A. Schirotzek, C. H. Schunck, and W.
Ketterle, Science, 2006, 311: 492

\bibitem{80.} J. Stenger, S. Inouye, A. P. Chikkatur, D. M.
Stamper-Kurn, D. E. Pritchard, and W. Ketterle, Phys. Rev. Lett.,
1999, 82: 4569

\bibitem{81.} F. D. M. Haldane, Phys. Rev. Lett., 1988, 61: 2015

\bibitem{82.} M. Onoda and N. Nagaosa, Phys. Rev. Lett., 2003, 90:
206601

\bibitem{83.} C. X. Liu, X. L. Qi, X. Dai, Z. Fang, and S. C. Zhang,
Phys. Rev. Lett., 2008, 101: 146802

\bibitem{84.} R. Yu, W. Zhang, H. J. Zhang, S. C. Zhang, X. Dai, and
Z. Fang, Science, 2010, 329: 61

\bibitem{85.} Z. Qiao, S. A. Yang, W. Feng, W. K. Tse, J. Ding, Y.
Yao, J. Wang, and Q. Niu, Phys. Rev. B, 2010, 82: 161414

\bibitem{86.} Y. Zhang and C. Zhang, Phys. Rev. B, 2011, 84: 085123

\bibitem{87.} H. Li, L. Sheng, and D. Y. Xing, Phys. Rev. B, 2011,
84: 035310

\bibitem{88.} L. B. Shao, S. L. Zhu, L. Sheng, D. Y. Xing, and Z. D.
Wang, Phys. Rev. Lett., 2008, 101: 246810

\bibitem{89.} C. Wu, Phys. Rev. Lett., 2008, 101: 186807

\bibitem{90.} N. Gemelke, Ph.D. thesis, Stanford University, 2007

\bibitem{91.} M. Zhang, Hsiang-hsuan Hung, C. Zhang, and C. Wu,
Phys. Rev. A, 2011, 83: 023615

\bibitem{92.} X. J. Liu, X. Liu, C. Wu, and J. Sinova, Phys. Rev. A,
2010, 81: 033622

\bibitem{93.} G. Juzeli\={u}nas and P. \"{O}hberg, Phys. Rev. Lett.,
2004, 93: 033602

\bibitem{94.} G. Juzeli\={u}nas, P. \"{O}hberg, J. Ruseckas, and A.
Klein, Phys. Rev. A, 2005, 71: 053614

\bibitem{95.} P. \"{O}hberg, G. Juzeli\={u}nas, J. Ruseckas, and M.
Fleischhauer, Phys. Rev. A, 2005, 72: 053632

\bibitem{96.} G. Juzeli\={u}nas, J. Ruseckas, P. \"{O}hberg, and M.
Fleischhauer, Phys. Rev. A, 2006, 73: 025602

\bibitem{97.} S. L. Zhu, H. Fu, C. J. Wu, S. C. Zhang, and L. M.
Duan, Phys. Rev. Lett., 2006, 97: 240401

\bibitem{98.} K. J. G\"{u}nter, M. Cheneau, T. Yefsah, S. P. Rath,
and J. Dalibard, Phys. Rev. A, 2009, 79: 011604(R)

\bibitem{99.} J. Dalibard, F. Gerbier, G. Juli\={u}nas, and P.
\"{O}hberg, arXiv: cond-mat/1008.5378

\bibitem{100.} J. Ruseckas, G. Juzeli\={u}nas, P. \"{O}hberg, and M.
Fleischhauer, Phys. Rev. Lett., 2005, 95: 010404

\bibitem{101.} F. Wilczek and A. Zee, Phys. Rev. Lett., 1984, 52:
2111

\bibitem{102.} G. Juzeli\={u}nas, J. Ruseckas, M. Lindberg, L.
Santos, and P. \"{O}hberg,  Phys. Rev. A, 2008, 77: 011802(R)

\bibitem{103.} S. L. Zhu, D. W. Zhang, and Z. D. Wang, Phys. Rev.
Lett., 2009, 102: 210403

\bibitem{104.} M. Merkl, F. E. Zimmer, and P. \"{O}hberg,  Europhys.
Lett., 2008, 83: 54002

\bibitem{105.} D. W. Zhang, Z. Y. Xue, H. Yan, Z. D. Wang, and S. L.
Zhu, arXiv: cond-mat/1104.0444

\bibitem{106.} N. R. Cooper, Adv. Phys., 2008, 57: 539

\bibitem{107.} M. Burrello and A. Trombettoni, Phys. Rev. Lett.,
2010, 105: 125304

\bibitem{108.} X. J. Liu, X. Liu, L. C. Kwek, C. H. Oh, Front.
Phys., 2008, 3(2): 113

\bibitem{109.} I. B. Spielman, Phys. Rev. A, 2009, 79: 063613

\bibitem{110.} Y. J. Lin, R. L. Compton, A. R. Perry, W. D.
Phillips, J. V. Porto, and I. B. Spielman, Phys. Rev. Lett., 2009,
102: 130401

\bibitem{111.} Y. J. Lin, R. L. Compton, K. Jim\'{e}nez-Garc\'{i}a,
J. V. Porto, and I. B. Spielman, Nature, 2009, 462: 628

\bibitem{112.} Y. J. Lin, R. L. Compton, K. Jim\'{e}nez-Garc\'{i}a,
W. D. Phillips, J. V. Porto, and I. B. Spielman, Nature Phys., 2011,
7: 531

\bibitem{113.} T. D. Stanescu, B. Anderson, and V. Galitski, Phys.
Rev. A, 2008, 78: 023616

\bibitem{114.} C. J. Wang, C. Gao, C. M. Jian and H. Zhai, Phys.
Rev. Lett., 2010, 105: 160403

\bibitem{115.} T. L. Ho and S. Zhang, arXiv: cond-mat/1007.0650

\bibitem{116.} Y. J. Lin, K. Jim\'{e}nez-Garc\'{i}a, and I. B.
Spielman, Nature, 2011, 471: 83

\bibitem{117.} X. J. Liu, M. F. Borunda, X. Liu and J. Sinova, Phys.
Rev. Lett., 2009, 102: 046402

\bibitem{118.} G. Juzeli\={u}nas, J. Ruseckas, and J. Dalibard,
Phys. Rev. A, 2010, 81: 053403

\bibitem{119.} Z. Lan, N. Goldman, A. Bermudez, W. Lu, and P.
\"{O}hberg, Phys. Rev. B, 2011, 84: 165115

\bibitem{120.} M. Merkl, A. Jacob, F. E. Zimmer, P. \"{O} hberg, and
L. Santos, Phys. Rev. Lett., 2010, 104: 073603

\bibitem{121.} L. H. Haddad and L. D. Carr, Physica D: Nonlinear
Phenomena, 2009, 238: 1413

\bibitem{122.} L. H. Haddad and L. D. Carr, Europhys. Lett., 2011,
94: 56002

\bibitem{123.} Z. Chen and B. Wu, Phys. Rev. Lett., 2011, 107:
065301

\bibitem{124.} See the review, W. Zawadzki and T. M Rusin, J. Phys.:
Condens. Matter, 2011, 23: 143201

\bibitem{125.} J. Schliemann, D. Loss, and R. M. Westervelt, Phys.
Rev. Lett., 2005, 94: 206801

\bibitem{126.} M. I. Katsnelson, Eur. Phys. J. B, 2006, 51: 157

\bibitem{127.} B. Trauzettel, Y. M. Blanter, and A. F. Morpurgo,
Phys. Rev. B, 2007, 75: 035305

\bibitem{128.} D. Luri\'{e} and S. Cremer, Physica, 1970, 50: 224

\bibitem{129.} X. Zhang, Phys. Rev. Lett., 2008, 100: 113903

\bibitem{130.} F. Dreisow, M. Heinrich, R. Keil, A. T\"{u}nnermann,
S. Nolte, S. Longhi, and A. Szameit, Phys. Rev. Lett., 2010, 105:
143902

\bibitem{131.} L. Lamata, J. Le\'{o}n, T. ScH\"{a}tz, and E. Solano,
Phys. Rev. Lett., 2007, 98: 253005

\bibitem{132.} J. Y. Vaishnav and C. W. Clark, Phys. Rev. Lett.,
2008, 100: 153002

\bibitem{133.} J. J. Song and B. A. Foreman, Phys. Rev. A, 2009, 80:
045602

\bibitem{134.} Q. Zhang, J. Gong, and C. H. Oh, Phys. Rev. A, 2010,
81, 023608

\bibitem{135.} D. Witthaut, Phys. Rev. A, 2010, 82: 033602

\bibitem{136.} J. Larson, J. P. Martikainen, A. Collin and E.
Sj\"{o}qvist, Phys. Rev. A, 2010, 82: 043620


\bibitem{137.} M. I. Katsnelson, K. S. Novoselov, and A. K. Geim,
Nature Phys., 2006, 2: 620

\bibitem{138.} O. Bahat-Treidel, O. Peleg, M. Grobman, N. Shapira,
M. Segev, and T. Pereg-Barnea, Phys. Rev. Lett., 2010, 104: 063901

\bibitem{139.} D. Witthaut, T. Salger, S. Kling, C. Grossert, and M.
Weitz, Phys. Rev. A, 2011, 84: 033601

\bibitem{140.} T. Salger, C. Grossert, S. Kling, and M. Weitz,
arXiv: cond-mat/1108.4447


\bibitem{141.} J. Otterbach, R. G. Unanyan, and M. Fleischhauer,
Phys. Rev. Lett., 2009, 102: 063602

\bibitem{142.}  M. G. Tarallo, J. Miller, J. Agresti, E. D'Ambrosio,
R. DeSalvo, D. Forest, B. Lagrange, J. M. Mackowsky, C. Michel, J.
L. Montorio , N. Morgado, L. Pinard, A. Remilleux, B. Simoni, P.
Willems, Appl. Opt., 2007, 46: 3348

\bibitem{143.} L. Khaykovich, F. Schreck, G. Ferrari, T. Bourdel, J.
Cubizolles, L. D. Carr, Y. Castin, and C. Salomon, Science, 2002,
296: 1290

\bibitem{144.} N. Dombey, P. Kennedy, and A. Calogeracos, Phys. Rev.
Lett., 2000, 85: 1787

\bibitem{145.} F. Wilczek, Nature Phys., 2009, 5: 614

\bibitem{146.} R. F. Service, Science, 2011, 332: 193

\bibitem{147.} S. L. Zhu, L. B. Shao, Z. D. Wang, and L. M. Duan,
Phys. Rev. Lett., 2011, 106: 100404

\bibitem{148.} L. Jiang, T. Kitagawa, J. Alicea, A. R. Akhmerov, D.
Pekker, G. Refael, J. I. Cirac, E. Demler, M. D. Lukin, and P.
Zoller, Phys. Rev. Lett., 2011, 106: 220402

\bibitem{149.} S. Tewari, S. Das. Sarma, C. Nayak, C. Zhang, and P.
Zoller, Phys. Rev. Lett., 2007, 98: 010506

\bibitem{150.} C. Nayak, S. H. Simon, A. Stern, M. Freedman, and S.
Das. Sarma, Rev. Mod. Phys., 2008, 80: 1083

\bibitem{151.} J. Alicea, Y. Oreg, G. Refael, F. von Oppen and M. P.
A. Fisher, Nature Phys., 2011, 7: 412

\bibitem{152.} J. Casanova, C. Sabin, J. Leon, I. L. Egusquiza, R.
Gerritsma, C. F. Roos, J. J. Garcia-Ripoll, E. Solano, arXiv:
1102.1651, 2011

\bibitem{153.} J. I. Cirac, P. Maraner, and J. K. Pachos, Phys. Rev.
Lett., 2010, 105: 190403

\end{thebibliography}
\end{document}